\documentclass[aps,prb,twocolumn,superscriptaddress,longbibliography]{revtex4-1}

\usepackage{multirow}
\usepackage{amsthm,amsmath,amsfonts}
\usepackage{comment}
\usepackage{longtable,tabularx}
\usepackage{subfigure,xcolor,color}
\usepackage{siunitx}
\usepackage{graphicx}
\usepackage{dcolumn}
\usepackage{float}
\usepackage{bbm,url}

\newtheorem{defn}{Definition}[section]

\newcommand{\Tr}{\mathrm{Tr}}

\newcommand{\bra}[1]{\mbox{$\langle #1 |$}}
\newcommand{\ket}[1]{\mbox{$| #1 \rangle$}}

\newcommand{\Ho}{\mathcal{H}^{(1)}}
\newcommand{\F}{\mathcal{F}}

\begin{document}
\author{Lexin Ding}
\affiliation{Faculty of Physics, Arnold Sommerfeld Centre for Theoretical Physics (ASC),\\Ludwig-Maximilians-Universit{\"a}t M{\"u}nchen, Theresienstr.~37, 80333 M{\"u}nchen, Germany}
\affiliation{Munich Center for Quantum Science and Technology (MCQST), Schellingstrasse 4, 80799 M{\"u}nchen, Germany}

\author{Sam Mardazad}
\affiliation{Faculty of Physics, Arnold Sommerfeld Centre for Theoretical Physics (ASC),\\Ludwig-Maximilians-Universit{\"a}t M{\"u}nchen, Theresienstr.~37, 80333 M{\"u}nchen, Germany}
\affiliation{Munich Center for Quantum Science and Technology (MCQST), Schellingstrasse 4, 80799 M{\"u}nchen, Germany}

\author{Sreetama Das}
\affiliation{Faculty of Physics, Arnold Sommerfeld Centre for Theoretical Physics (ASC),\\Ludwig-Maximilians-Universit{\"a}t M{\"u}nchen, Theresienstr.~37, 80333 M{\"u}nchen, Germany}
\affiliation{Munich Center for Quantum Science and Technology (MCQST), Schellingstrasse 4, 80799 M{\"u}nchen, Germany}

\author{Szil\'ard Szalay}
\affiliation{Strongly Correlated Systems Lend{\"u}let Research Group, Wigner Research Centre for Physics, 29-33, Konkoly-Thege Mikl\'os str., H-1121,  Budapest, Hungary}

\author{Ulrich Schollw{\"o}ck}
\affiliation{Faculty of Physics, Arnold Sommerfeld Centre for Theoretical Physics (ASC),\\Ludwig-Maximilians-Universit{\"a}t M{\"u}nchen, Theresienstr.~37, 80333 M{\"u}nchen, Germany}
\affiliation{Munich Center for Quantum Science and Technology (MCQST), Schellingstrasse 4, 80799 M{\"u}nchen, Germany}

\author{Zolt\'an Zimbor{\'a}s}
\affiliation{Theoretical Physics Department, Wigner Research Centre for Physics, P.O.Box 49 H-1525, Budapest, Hungary}
\affiliation{MTA-BME Lend\"ulet Quantum Information Theory Research Group, Budapest, Hungary}
\affiliation{Mathematical Institute, Budapest University of Technology and Economics, P.O.Box 91 H-1111, Budapest, Hungary}

\author{Christian Schilling}
\email{c.schilling@physik.uni-muenchen.de}
\affiliation{Faculty of Physics, Arnold Sommerfeld Centre for Theoretical Physics (ASC),\\Ludwig-Maximilians-Universit{\"a}t M{\"u}nchen, Theresienstr.~37, 80333 M{\"u}nchen, Germany}
\affiliation{Munich Center for Quantum Science and Technology (MCQST), Schellingstrasse 4, 80799 M{\"u}nchen, Germany}
\affiliation{Wolfson College, University of Oxford, Linton Rd, Oxford OX2 6UD, United Kingdom}

\title{Concept of orbital entanglement and correlation in quantum chemistry}

\begin{abstract}
A recent development in quantum chemistry has established the quantum mutual information between orbitals as a major descriptor of electronic structure. This has already facilitated remarkable improvements of numerical methods and may lead to a more comprehensive foundation for chemical bonding theory. Building on this promising development, our work provides a refined discussion of quantum information theoretical concepts by introducing the physical correlation and its separation into classical and quantum parts as distinctive quantifiers of electronic structure. In particular, we succeed in quantifying the entanglement. Intriguingly, our results for different molecules reveal that the total correlation between orbitals is mainly classical, raising questions about the general significance of entanglement in chemical bonding.
Our work also shows that implementing the fundamental particle number superselection rule, so far not accounted for in quantum chemistry, removes a major part of correlation and entanglement previously seen. In that respect, realizing quantum information processing tasks with molecular systems might be more challenging than anticipated.
\end{abstract}

\maketitle

\section{Introduction}\label{sec:intro}
Quantum information theory has had a strong impact in many-body quantum physics, particularly quantum chemistry and solid state physics in recent years. For instance, the concepts of entanglement and correlation became important descriptors of distinctive behavior and features of quantum systems.\cite{amico2008entanglement,EisertReview} Prime examples are quantum phase transitions\cite{Osborne02,Fazio02,Vidal03, calabrese2004entanglement} and topological order.\cite{Haldane08, kitaev2006topological,levin2006detecting} Moreover, the reduced entanglement due to the locality of lattice models in solid state physics has manifested itself in a highly efficient ground state approach known as the density matrix renormalization group ansatz.\cite{White1992,Schollwoeck2011} The recent success of this approach in quantum chemistry,\cite{White99,White00,Fano01,Chan11,Neck14,Legeza15Rev} however, seems to be rather surprising: Why shall a molecular Hamiltonian expressed in second quantization exhibit any form of locality on the artificial one-dimensional lattice formed by molecular orbitals? Certainly, this success can partly be traced back to the efficiency of the underlying tensor network ansatz.\cite{Neck14} Yet, the identification of optimal reference orbitals to restore at least some form of locality is fundamentally important.\cite{White05,Chan08} A sophisticated procedure for determining those reference orbitals has recently been developed with a remarkable increase of numerical efficiency.\cite{Reiher11,Eisert16mode} To be more specific, this orbital transformation improves successively the choice of orbitals $\{\varphi_j\}$ by analyzing their pairwise correlation in terms of the quantum mutual information
\begin{equation}\label{mutI}
I_{i,j} = S(\rho_i) + S(\rho_j) - S(\rho_{ij})
\end{equation}
to increase the locality of the respective ``lattice'' model. Indeed, $I_{i,j}$ quantifies the \emph{additional} quantum information content of the two orbital reduced density matrix $\rho_{ij}$ beyond the one of the single-orbital density matrices $\rho_i,\rho_j$, where $S(\rho) \equiv -\mbox{Tr}[\rho\ln{(\rho)}]$ is the von Neumann entropy.\cite{lindblad1973entropy} In the same context, this tool \eqref{mutI} has also been used to automate the selection of active orbital spaces.\cite{Legeza03,Reiher16,Reiher17a} Hence, the significance of the quantum mutual information for the development of a successful density matrix renormalization group method in quantum chemistry can hardly be overestimated.

Besides its fruitful utilization in numerical methods, the quantum mutual information \eqref{mutI} also has been put forward as a tool for describing and quantifying the bonding structure of molecular systems.\cite{Reiher11,Reiher12,Reiher13,Reiher14,Reiher15,Ayers15a,Ayers15b,Bogus15,Szalay17,Reiher17b,Legeza18,Legeza19}
Its information theoretical meaning can thus be seen as a temptation to add a quantum information theoretical facet to this ongoing development. Quantum effects certainly are central in chemical bonding but does this already confirm the distinctive role of entanglement in quantum chemistry? The recent works \cite{Reiher11,Reiher12,Reiher13,Reiher14,Reiher15,Ayers15a,Ayers15b,Bogus15,Szalay17,Reiher17b,Legeza18,Legeza19} seem to suggest an affirmative answer by referring in their analyses extensively to ``orbital entanglement''. Yet, the quantum mutual information does
not quantify the entanglement since it includes both quantum \emph{and} classical correlation.\cite{henderson2000information,henderson2001classical,groisman2005quantum}. In more recent works such as Ref.~\onlinecite{Legeza18} this has been understood as well. Yet, no further work has been done along those lines and the term entanglement is still often used in the literature as an incorrect synonym for total correlation. As a matter of fact it also does not quantify the true physical correlation but instead overestimates it\cite{Wiseman03,bartlett2003entanglement,wise04fluffy,banuls2007entanglement,banuls2009entanglement,Friis13,eisler2015partial,Friis16,spee2018mode} by taking into account the so-called ``fluffy bunny''-correlation.\cite{wise04fluffy} The latter emerges as a mathematical artifact from the choice of incorrect algebras of observables.\cite{SSR,wick1970superselection, wick1997intrinsic,Lexin20a}

The aim of our work is to provide a framework for addressing those open challenges based on a refined discussion of quantum information theoretical concepts: We introduce the physical correlation and its classical and quantum parts as distinctive quantifiers of electronic structure. In particular, we describe their relations to the quantum mutual information \eqref{mutI}. In that context, we justify the  particle number superselection rule, highlight its significance and derive the required modification of \eqref{mutI}. Our work is thus building on the recent development,\cite{Reiher11,Reiher12,Reiher13,Reiher14,Reiher15,Ayers15a,Ayers15b,Bogus15,Reiher16,Reiher17a,Szalay17,Reiher17b,Legeza18,Legeza19}
also by strengthening its quantum information theoretical facet. In that respect, we would like to recall that entanglement is an
important resource for realizing quantum cryptography,\cite{Ekert91,Zeilinger00} superdense coding\cite{bennett1992communication,fang2000experimental,ye2005scheme,schaetz2004quantum}  and possibly even quantum computing.\cite{Jozsa03}
Those applications clearly necessitate quantification of entanglement and total correlation in an \emph{operationally meaningful} way.

Our comprehensive analysis of molecular systems eventually leads to two key results. First, taking into account the fundamental particle number superselection rule removes a major part of the correlation and entanglement previously seen between molecular orbitals. Second, the remaining correlation is mainly classical. Hence, the role of entanglement in quantum chemistry needs to be reassessed and realizing quantum information processing tasks with molecular systems might be more challenging than expected.

The paper is structured as follows. In Section \ref{sec:QIT} we briefly recall the quantum information theoretical concepts of correlation and entanglement, introduce and prove the particle number superselection rule and explain how it affects entanglement and correlation.
Section \ref{sec:analytic} provides an analytic illustration of those concepts in the form of elementary examples.
In Section \ref{sec:results} we quantify the orbital correlation and its separation into quantum and classical parts for the water, naphthalene and dichromium molecules by exact numerical means.

\section{Quantum Information Theoretical Concepts}\label{sec:QIT}
This section is devoted to covering some important concepts in quantum information theory that have great relevance in orbital correlation and entanglement theory. First, we present the definitions and measures for different notions of correlation for distinguishable systems. Next, we show how the anticommutation property of  fermionic operators necessitates the introduction of superselection rules, and then describe how these rules affect what the physically accessible correlation and entanglement is in a quantum state. Taking this into consideration, we provide the operationally meaningful correlation and entanglement measures that will be used throughout the paper.


\subsection{Total correlation, entanglement and classical correlation}\label{sec:IEC}
Let $\mathcal{H}$ be a finite-dimensional Hilbert space. A quantum state can be represented by a density operator $\rho$ acting on $\mathcal{H}$, which satisfies positivity (non-negative eigenvalues) and trace unity ($\Tr[\rho]=1$). Together they form the convex set $\mathcal{D}$ of all density operators, with the extreme points being the pure states, $\rho \equiv \ket{\Psi}\bra{\Psi}$. If the system consists of two distinct subsystems, handled by Alice and Bob, the Hilbert space can be split into a tensor product, $\mathcal{H} = \mathcal{H}_A \otimes \mathcal{H}_B$. In this case all Hermitian operators on each subsystem are physical observables. The algebra of observables on the total system can be written as the tensor product $\mathcal{B}(\mathcal{H}) = \mathcal{B}(\mathcal{H}_A) \otimes \mathcal{B}(\mathcal{H}_B)$. For a physical observable $\mathcal{O}$ acting on $\mathcal{H}$, its expectation value with respect to a state $\rho$ can be calculated as
\begin{equation}
\langle \mathcal{O} \rangle_\rho = \Tr[\rho \,\mathcal{O}].
\end{equation}
If Alice wants 
 to perform a local measurement corresponding to $\mathcal{O}_A \in \mathcal{B}(\mathcal{H}_A)$, the expectation value of the measurement outcome is
\begin{equation}
\langle \mathcal{O}_A \otimes \mathbbm{1}_B \rangle_{\rho} = \Tr[\rho(\mathcal{O}_A \otimes \mathbbm{1}_B)]. \label{eqn:local_mes}
\end{equation}
Since such a measurement is essentially restricted to only Alice's subsystem, there exists a local description of the quantum state, with respect to which the expectation of any local operator $\mathcal{O}_A$ is the same as \eqref{eqn:local_mes}. This is the commonly used reduced density matrix defined as
\begin{equation}
\rho_A \equiv \Tr_B[\rho],
\end{equation}
which satisfies
\begin{equation}
\langle \mathcal{O}_A \rangle_{\rho_A} = \langle \mathcal{O}_A \otimes \mathbbm{1}_B \rangle_\rho, \quad \forall \mathcal{O}_A \in \mathcal{B}(\mathcal{H}_A).
\end{equation}

\begin{figure}[!t]
    \centering
    \includegraphics[scale=0.25]{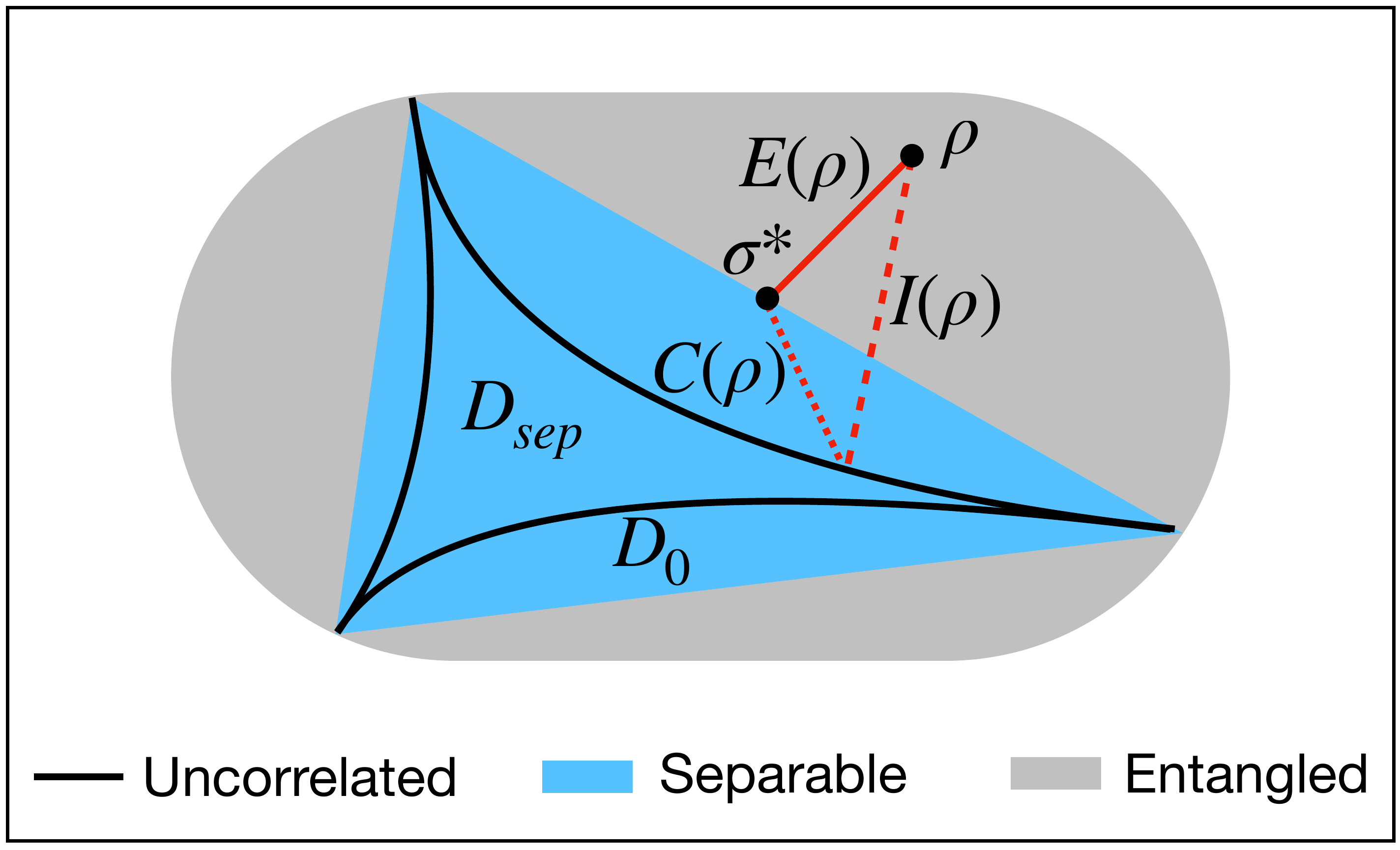}
    \caption{Schematic illustration of the space of quantum states, including the uncorrelated ($\mathcal{D}_0$, black curve) and separable states ($\mathcal{D}_{sep}$, blue). The total correlation $I$ (red dashed) and entanglement $E$ (red solid) of a state $\rho$ is its distances to $\mathcal{D}_0$ and $\mathcal{D}_{sep}$, respectively, measured by the relative entropy. The classical correlation $C$ is the distance from the closest separable state $\sigma^\ast$ to the closest product state (red dotted).}
    \label{fig:states}
\end{figure}

The notion of correlation, classical or quantum mechanical, has deep connection to the notion of local measurements.\cite{horodecki2009quantum} For a pair of observables $\mathcal{O}_A$ and $\mathcal{O}_B$ acting on subsystems $A$ and $B$, respectively, one can compare the outcomes of separate or joint measurements, and infer some aspects of correlation. For example, the correlation function
\begin{equation}
    C_\rho(\mathcal{O}_A, \mathcal{O}_B) \equiv \langle \mathcal{O}_A \otimes \mathcal{O}_B \rangle_\rho - \langle \mathcal{O}_A \rangle_{\rho_A} \langle \mathcal{O}_B \rangle_{\rho_B} \label{eqn:corrfn}
\end{equation}
defined as the difference of expectation values between joint and separate measurement, quantifies the correlation in a state $\rho$ with respect to a given pair of observables. The set of uncorrelated states is defined as those states that are uncorrelated with respect to \textit{any} pair of observables $\mathcal{O}_A \in \mathcal{B}(\mathcal{H}_A)$ and $\mathcal{O}_B\in \mathcal{B}(\mathcal{H}_B)$:
\begin{defn} \label{def:uncorr}
A density operator $\rho$ represents an uncorrelated state if and only if
\begin{equation}
    \langle \mathcal{O}_A \otimes \mathcal{O}_B \rangle_\rho = \langle \mathcal{O}_A \rangle_{\rho_A} \langle \mathcal{O}_B \rangle_{\rho_B},
\end{equation}
for \textit{any} pair of operators $\mathcal{O}_A$ and $\mathcal{O}_B$ acting on subsystems A and B, respectively. The set of uncorrelated states is denoted by $\mathcal{D}_0$. A state $\rho$ is called correlated if $\rho \notin \mathcal{D}_0$.
\end{defn}
For a {\it distinguishable} bipartite system, Definition \ref{def:uncorr} is equivalent to stating that an uncorrelated state $\rho$ can be written as a product state
\begin{equation}
    \rho = \rho_A \otimes \rho_B. \label{eqn:productstates}
\end{equation}
Eq.~\eqref{eqn:productstates} further  implies that the uncorrelated states are the ones that can be prepared locally by Alice and Bob. If Alice and Bob can also communicate classically, they can also prepare correlated states as probabilistic distributions of uncorrelated states, namely mixtures of product states. The correlation contained in those states is purely classical.\cite{werner1989quantum}  Therefore the set of such \emph{separable} states is the same as all convex combinations of elements from $\mathcal{D}_0$ (see also Figure \ref{fig:states} for an illustration):
\begin{defn}\label{def:separable}
A density operator $\rho$ represents a separable state if and only if $\rho \in \mathcal{D}_{sep} \equiv\mathrm{Conv}(\mathcal{D}_{0})$, where $\mathrm{Conv}(\mathcal{D}_{0})$ denotes the convex hull of $\mathcal{D}_{0}$.  A state $\rho$ called entangled if and only if $\rho \notin \mathcal{D}_{sep}$.
\end{defn}
For \emph{pure} states, the entanglement entropy is a (unique) entanglement measure,\cite{horodecki2009quantum, bennett1996concentrating} defined as
\begin{equation}
    E(\ket{\Psi}\bra{\Psi}) = S(\rho_{A/B}), \label{eqn:ententr}
\end{equation}
where $S$ is the von Neumann entropy
\begin{equation}
    S(\rho) = -\Tr[\rho \ln(\rho)].
\end{equation}
Notice that the reduced density matrix in \eqref{eqn:ententr} can refer to either of the subsystems. This is due to the existence of Schmidt decompositions in bipartite pure states, and consequently $\rho_A$ and $\rho_B$ are isospectral. Unfortunately this is no longer true if the total state is a mixed state, and \eqref{eqn:ententr} is not a valid entanglement measure anymore. This is obvious as one could simply consider mixed product states $\rho_{A} \otimes \rho_{B}$, which are uncorrelated and thus also separable even if $ S(\rho_A) \ne 0$. However, there is a widely accepted \textit{total correlation} measure for general mixed states,\cite{vedral1997quantifying,lindblad1973entropy,henderson2001classical,vedral2002role} namely the quantum mutual information
\begin{equation}
    I(\rho) \equiv S(\rho||\rho_A \otimes \rho_B) = S(\rho_A) + S(\rho_B)-S(\rho), \label{eqn:mutual_info}
\end{equation}
where $S(\cdot || \cdot)$ is the quantum relative entropy defined as
\begin{equation}
    S(\rho||\sigma) \equiv \Tr[\rho(\ln(\rho)-\ln(\sigma))].
\end{equation}
The quantity \eqref{eqn:mutual_info} measures the amount of information in the total state $\rho$ that is not yet contained in the reduced states $\rho_A, \rho_B$.\cite{horodecki2005local} Interestingly, for distinguishable bipartite systems, \eqref{eqn:mutual_info} can be rewritten as
\begin{equation}\label{eqn:mutImin}
    I(\rho) = \min_{\sigma \in \mathcal{D}_0} S(\rho || \sigma),
\end{equation}
where $\mathcal{D}_0$ is the set of uncorrelated states in Definition \ref{def:uncorr}. That is, the mutual information of a state $\rho$ is its  minimal distance to the set of uncorrelated states, measured by the quantum relative entropy. This appealing geometric interpretation of \eqref{eqn:mutual_info} is also illustrated in Figure \ref{fig:states}. The geometric picture of quantum states leads in the same fashion to an entanglement measure which is valid for mixed states as well. This \textit{relative entropy of entanglement} of a state $\rho$ is defined as its minimal distance to the set of \textit{separable} states\cite{vedral1997quantifying} measured in terms of the quantum relative entropy,
\begin{equation}
    E(\rho) = \min_{\sigma \in \mathcal{D}_{sep}} S(\rho||\sigma) = S(\rho||\sigma^\ast), \label{eqn:rel_ent}
\end{equation}
where $\sigma^\ast$ denotes the closest separable state to $\rho$, as illustrated in Figure \ref{fig:states}. This measure for entanglement has remarkable properties. First, the relative entropy of entanglement of a pure state coincides with the von Neumann entropy defined in \eqref{eqn:ententr}.\cite{henderson2000information} Second, it quantifies in a pure state the number of standard units of entanglement (``Bell pairs'') one can extract.\cite{bennett1996concentrating} Last but not least, as an entropic measure that fits together perfectly with the mutual information, it allows us to harness the rich information theoretical meaning of entanglement.

Knowing the closest separable state\footnote{In case the closest separable state $\sigma^\ast$ is not unique, we choose the one that results in the lowest classical correlation.}, the \emph{classical correlation} of $\rho$ can be quantified geometrically, namely as the distance from the closest separable state $\sigma^\ast$ to the closest uncorrelated state $\rho_A \otimes \rho_B$,\cite{henderson2001classical}
\begin{equation}\label{eq:classical}
C(\rho) \equiv S(\sigma^\ast||\rho_A \otimes \rho_B).
\end{equation}
The geometric picture of quantum states as well as the measures for total correlation, entanglement and classical correlation are illustrated in Figure \ref{fig:states}.

A comment is in order concerning the relation between the different types of correlation. In general, entanglement and classical correlation do not sum to the total correlation. This is because
mixed quantum states typically contain quantum correlations beyond entanglement as it is concisely described by the concept of quantum discord.\cite{Adesso16} Moreover, a known exact relation including quantum discord refers to an alternative definition of classical correlation which is more technical than our simple geometric one (see Eq.~\eqref{eq:classical}).\cite{Adesso16}
Due to its particular significance for quantum information tasks and for the sake of simplicity we focus in our work on entanglement, however, and refer to it occasionally as quantum correlation.

All these well-defined quantum information theoretical concepts can be established in the context of \emph{indistinguishable} particles as well, yet with some caveats. For a respective detailed discussion we refer the reader to Ref.~\onlinecite{Lexin20a}. To follow here the most natural route for establishing a notion of entanglement and correlation in fermionic quantum systems we consider the fermionic Fock space $\mathcal{F}[\Ho]$ with a $D$-dimensional one-particle Hilbert space $\Ho$. In the context of quantum chemistry, $\Ho$ is typically defined by choosing a finite basis set of spin-orbitals.  To establish a notation of subsystems and locality we split $\mathcal{H}_1$ into two (or more) complementary subspaces,
\begin{equation}\label{H1split}
\Ho = \Ho_A \oplus \Ho_B\,.
\end{equation}
For instance, this can be achieved by dividing the reference basis $\mathcal{B}=\{\ket{\varphi_j}\}_{j=1}^D$ of $\Ho$ into two disjoint subsets
$\mathcal{B}_A=\{\ket{\varphi_j}\}_{j=1}^{D_A}$, $\mathcal{B}_B=\{\ket{\varphi_j}\}_{j=D_A+1}^{D}$ of spin-orbitals.
By referring to second quantization with creation/annihiltion operators $f_j^{(\dagger)}$, vacuum state $\ket{\Omega}$
and introducing the configuration states
\begin{equation}\label{config}
\ket{n_1,\ldots,n_D} \equiv \big(f_1^\dagger\big)^{n_1}\ldots \big(f_D^\dagger\big)^{n_D} \ket{\Omega}
\end{equation}
the presence of a tensor product structure in the total Fock space $\F$ becomes apparent. Indeed, the splitting
\begin{eqnarray}\label{configsplit}
&&\ket{n_1,\ldots,n_{D_A},n_{D_A+1},\ldots,n_D} \mapsto \nonumber \\
&&\qquad  \ket{n_1,\ldots,n_{D_A}} \otimes  \ket{n_{D_A+1},\ldots,n_D}
\end{eqnarray}
establishes a decomposition $\F = \F_A \otimes \F_B$ and in that sense gives rise to a notion of so-called orbital or mode entanglement and correlation. It is worth noticing that calculating the orbital reduced density operator $\rho_A \equiv \Tr_B[\rho]$
means to reduce out spin-orbitals rather than particles. As a consequence, $\rho_A$ is a density matrix with indefinite particle number ``living'' on the Fock space $\F_A$ involving spin-orbitals $\mathcal{B}_A$. Actually, a few subtleties arise due to the fermionic anticommutation relations which are concerned with the definition of local algebras and reduced density matrices. They disappear, however, if one takes into account the parity superselection rule.\cite{Friis16,SzilardQIT} This fundamental rule of nature and its impact on the notion of correlation and entanglement are explained in the following.

\subsection{Significance of the superselection rule}\label{sec:SSR}
A key ingredient in the physics of fermionic systems is the so-called  {\it parity superselection rule} (P-SSR). In its original form,  P-SSR forbids superpositions of even and odd fermion-numbers states. In a more modern version, P-SSR states that the operators belonging to physically measurable quantities must commute with the particle parity operator. This means they have to be linear combinations of even degree monomials of the creation and annihilation operators. This in turn implies that a superposition of two pure states with even and odd particle numbers cannot be distinguished from  an incoherent classical mixture of those states, thus one recovers the original formulation as a consequence.

\begin{figure}[ht]
\includegraphics[scale=0.32]{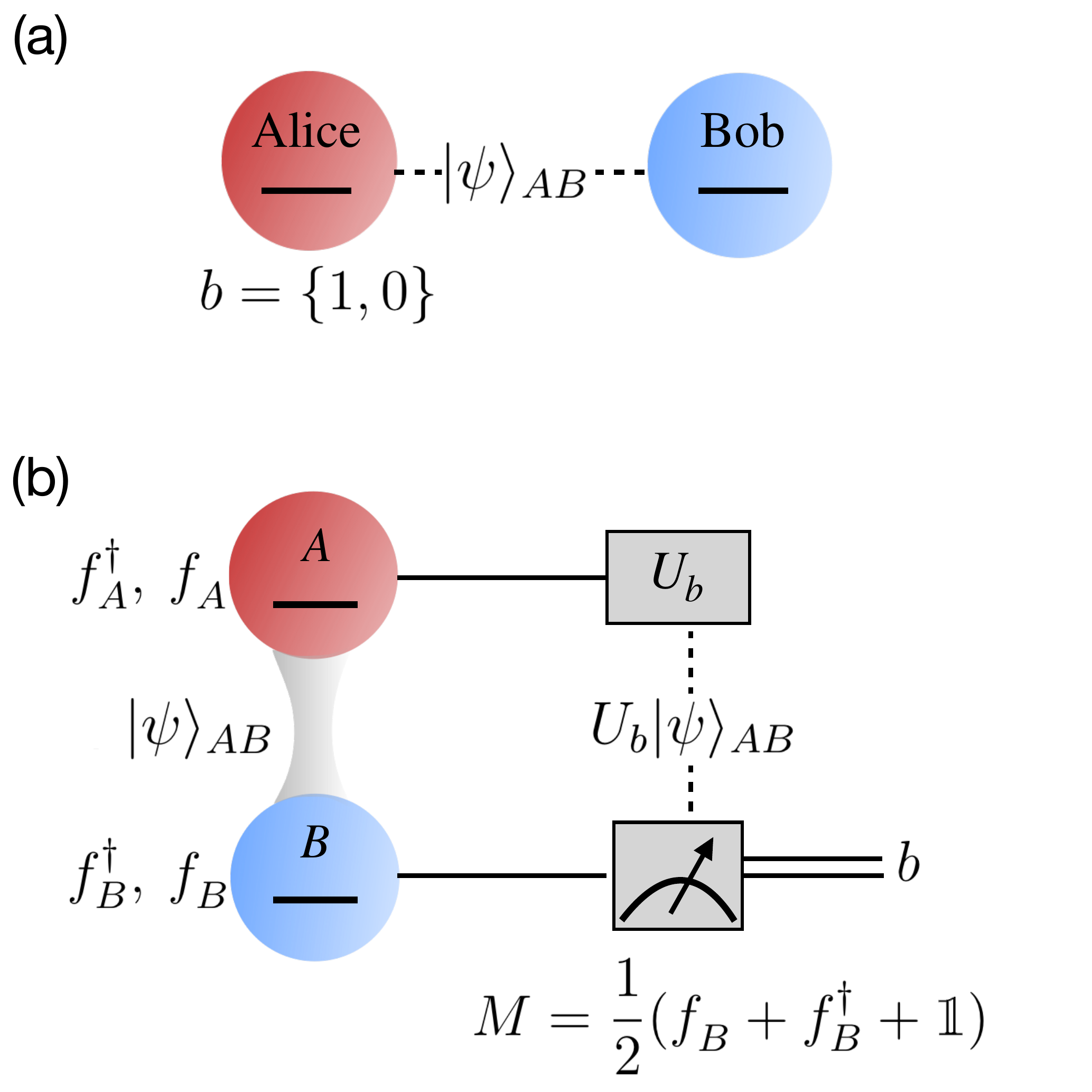}
\caption{(a) Two spatially separated parties Alice and Bob share a quantum state $|\psi\rangle_{AB}$. (b) The protocol showing how superluminal signaling is possible when parity superselection rule is broken: Alice communicates the bit value $b \in \{0,1 \}$ by applying the corresponding unitary $U_{b}$, Bob measures the observable $M$ and obtains instantaneously that bit value, as explained in the text.
}
\label{nosignalling}
\end{figure}

The idea that the laws of nature impose P-SSR on fermionic systems was originally derived based on group theoretical arguments.\cite{SSR,wick1970superselection, wick1997intrinsic} However, the pertinence of P-SSR is also obvious from the fundamental fact that violation of P-SSR would lead to a violation of the no-signaling theorem, as we will explain in the following.  The no-signaling theorem states that two spatially separated parties cannot communicate  faster than the speed of light. To relate this to the P-SSR, let us assume that two distant parties Alice and Bob could violate the P-SSR. For our argument it is sufficient for Alice and Bob to have each access to one mode (e.g., an atomic spin-orbital).  Their local Fock spaces are thus generated by the fermionic annihilation and creation operators $(f_{A},f_{A}^{\dagger})$ and $(f_{B},f_{B}^{\dagger})$, respectively. Assume now that they can share the state $|\psi\rangle_{AB} = \frac{1}{\sqrt{2}}(|0\rangle_{A}|0\rangle_{B} + |0\rangle_{A}|1\rangle_{B})$, which is a superposition of odd and even number states.
The procedure for Alice to communicate instantaneously one bit $b=0,1$ of classical information to Bob would be the following (see also Figure \ref{nosignalling}): both of them synchronize the clocks in their labs, and they pre-decide to perform local operations at a particular time. If Alice wants to communicate $1$, she does nothing (i.e., formally applies the unitary $U_1=\mathbbm{1}$), so $|\psi\rangle_{AB}$ remains unchanged; to communicate $0$, Alice applies the unitary  $U_0=i(f^{\dagger}_{A} - f_{A}^{\phantom{\dagger}})$, and the state becomes
$|\psi^{\prime}\rangle_{AB} =\frac{i}{\sqrt{2}}(|1\rangle_{A}|0\rangle_{B} + |1\rangle_{A}|1\rangle_{B})$. At the same instant Bob measures the observable $\frac{1}{2}(f_{B} + f_{B}^{\dagger}+\mathbbm{1})$. One easily verifies that in both cases $b=0,1$ the outcome of that measurement is deterministic and will be nothing else than the value of $b$. Hence, this proposed procedure allows Alice to communicate instantaneously a bit $b$ of information in contradiction to the no-signaling theorem and the laws of relativity.

Beside the parity superselection rule, it is often pertinent to consider superselection rules due to some experimental limitations. One such rule is the fermion {\it particle number superselection rule} (N-SSR). Measurable quantities obeying N-SSR must commute with the particle parity operators.\cite{wick1970superselection} This, in the form of lepton number conservation, was once considered to be an exact symmetry of Nature. Recently, however, there have been indications that fundamental Majorana particles may exist which could lead to a violation of the N-SSR. Nevertheless, in a usual quantum chemistry set-up, we can safely regard N-SSR to hold.
Indeed, the energies of common molecular systems are (in contrast to systems studied in high energy physics) sufficiently low to entirely suppress the emergence of electrons and other particles from vacuum fluctuations.
In the following parts of the paper we will in particular discuss the consequences of \emph{both} superselection rules, but we assume that the N-SSR is the more relevant one in quantum chemistry.

\subsection{Taking into account the superselection rule}\label{sec:SSRincorp}
Having established the fundamental importance of superselection rules, we will now elucidate how they affect our description of quantum states, and consequently change the physically accessible correlation and entanglement in a quantum state. Accordingly, the SSRs will have important consequences for the realization of quantum information processing tasks (e.g., for teleportation\cite{Potts20,Friis20}, quantum control\cite{zimboras2014dynamic}  and molecular error-correcting codes\cite{albert2020robust}).

In a word, SSRs are restrictions on local algebras of observables, resulting in physical algebras $\mathcal{A}_A$ and $\mathcal{A}_B$. If the SSR is related to some locally conserved quantity $Q_{A/B}$, then local operators must also preserve this quantity. That is, all local observables satisfy
\begin{equation}
\mathcal{A}_{A/B} \ni \mathcal{O}_{A/B} = \sum_q P_q \mathcal{O}_{A/B} P_q,
\end{equation}
where $q$ ranges over all possible value of $Q_{A/B}$ and $P_q$'s are projectors onto the eigensubspaces, i.e. $\mathcal{O}_{A/B}$ are block diagonal in any eigenbasis of $Q_{A/B}$.  It follows that different SSRs will lead to drastically different $\mathcal{A}_{A/B}$. The fact that we cannot physically implement every mathematical operator changes the accessibility of quantum states. The fully accessible states are called the physical states, and they satisfy
\begin{equation}
\rho = \sum_{q,q'} P_q \otimes P_{q'} \rho P_q \otimes P_{q'},
\end{equation}
or equivalently
\begin{equation}
[\rho, Q_{A/B}] = 0. \label{eqn:SSRcom}
\end{equation}
For a general state $\rho$ which does not satisfy \eqref{eqn:SSRcom}, we can obtain its physical part by the following projection
\begin{equation}
\rho^\textrm{Q} \equiv  \sum_{q,q'} P_q \otimes P_{q'} \rho P_q \otimes P_{q'}. \label{eqn:tilde}
\end{equation}
The physical state $\rho^\textrm{Q}$ gives the same expectation value as $\rho$ for all \textit{physical} observables. Therefore we can define a new class of uncorrelated states to be the ones with uncorrelated physical parts with respect to the physical algebra:
\begin{equation}
\begin{split}
\mathcal{D}_0^{\textrm{Q-SSR}} = \{ \rho \, | \,& \langle \mathcal{O}_A \otimes \mathcal{O}_B \rangle_{\rho} =   \langle \mathcal{O}_A  \rangle_{\rho_A}  \langle\mathcal{O}_B \rangle_{\rho_B},
\\
&\forall \mathcal{O}_A \in \mathcal{A}_A, \mathcal{O}_B \in \mathcal{A}_B \}.
\end{split}
\end{equation}
It is clear that the new set of uncorrelated states includes the one of the distinguishable setting, i.e. $\mathcal{D}_0 \subseteq \mathcal{D}^{\textrm{Q-SSR}}_0$. Consequently also more states are deemed separable. Relating to Figure \ref{fig:states}, both the correlation and entanglement measure become smaller in the presence of an SSR. There are two key messages here. First of all, correlation and entanglement are relative concepts. They depend not only on the particular division of the total system into two (or more) subsystems but also on the underlying SSRs, which eventually defines the physical local algebras of observables $\mathcal{A}_{A/B}$ and the global algebra $\mathcal{A}_A \otimes \mathcal{A}_B$. Secondly, by ignoring the fundamentally important SSRs, one may radically overestimate the amount of physical correlation and entanglement in a quantum state.

\begin{figure}[!t]
    \centering
    \includegraphics[scale=0.4]{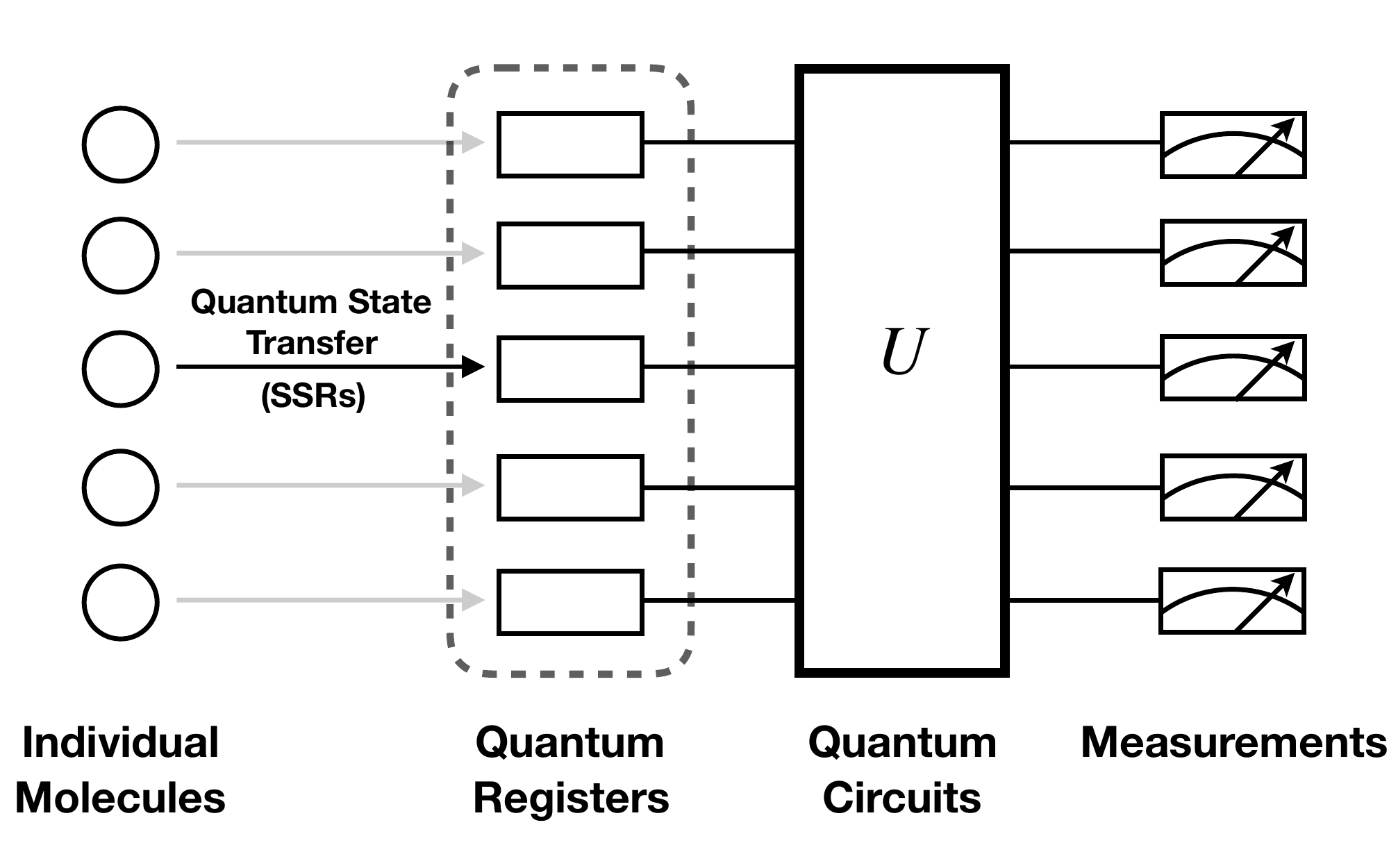}
    \caption{Schematic protocol for utilizing entanglement from molecular systems (see text for more details).}
    \label{fig:QCScheme}
\end{figure}

One of the biggest motivation for correctly identifying the amount of physical correlation and entanglement in a quantum state is its value for information processing tasks. An operationally meaningful quantification of entanglement does not only reveal non-local properties of a quantum state, but should also measure the amount of resource that can be extracted for performing various quantum information tasks mentioned in Section \ref{sec:intro}. In Figure \ref{fig:QCScheme} we illustrate the schematic protocol for utilizing entanglement from molecular systems. The quantum states of individual molecules are transferred to SSR-free quantum registers with Hilbert spaces of equal or higher dimensions, through local measurements and classical communication. A quantum circuit represented by a unitary gate $U$ in Figure \ref{fig:QCScheme} then acts on these quantum register states to perform computations. Finally, the end results of the computation are retrieved with carefully designed measurements. The key step that limits the extraction of entanglement is the transferring of the quantum state, which is constrained by the underlying  SSR\cite{bartlett2003entanglement}. What remains on the quantum registers after the transfer are the \emph{physical} parts defined in Eq.~\eqref{eqn:tilde}. From this perspective, the Q-SSR-constrained total correlation, entanglement and classical correlation of a single system in a state $\rho$ follow as
\begin{equation}
\begin{split}
I^{\textrm{Q-SSR}}(\rho) &= I(\rho^\textrm{Q}),
\\
E^{\textrm{Q-SSR}}(\rho) &= E(\rho^\textrm{Q}),
\\
C^{\textrm{Q-SSR}}(\rho) &= C(\rho^{\textrm{Q}}). \label{eqn:SSRmeasures}
\end{split}
\end{equation}

\begin{figure}[!t]
\centering
\includegraphics[scale=0.291]{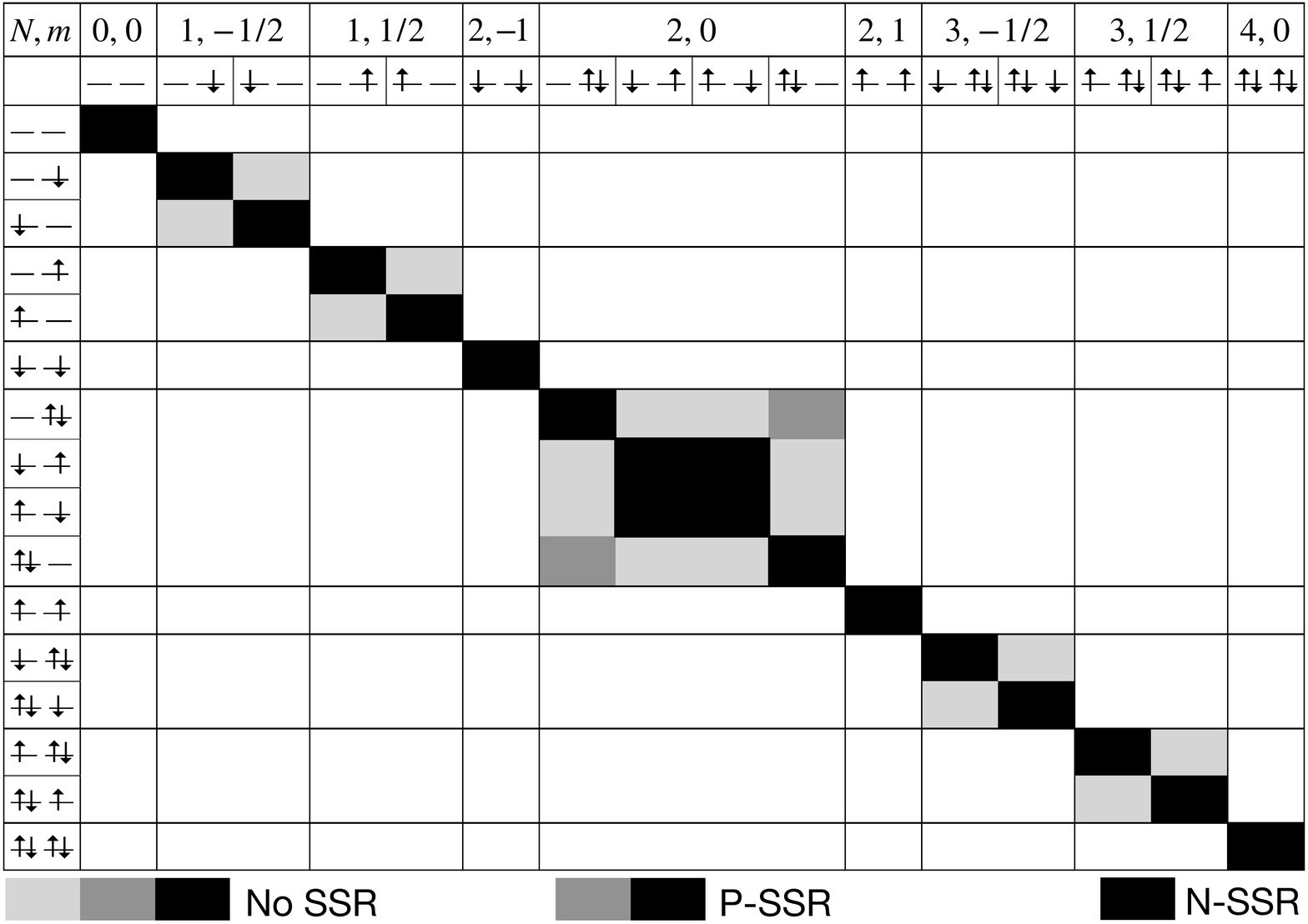}
\caption{Illustration of two-orbital reduced density matrix $\rho_{i,j}$. Most entries vanish due spin and particle symmetry (white). According to \eqref{eqn:tilde} the P-SSR sets light gray entries to zero while N-SSR removes in addition two entries (gray).}
\label{fig:sectors}
\end{figure}

All quantum information theoretical concepts discussed so far are applicable to any \emph{arbitrary} orthonormal basis of $D\geq 2$ spin orbitals. In particular, they then refer to any \emph{arbitrary} separation of them into subsystems $A$ and $B$ defined by spin-orbitals $\{\ket{\varphi_j}\}_{j=1}^{D_A}$ and $\{\ket{\varphi_j}\}_{j=D_A+1}^D$, respectively, according to \eqref{H1split}.
As far as the description of electronic structure is concerned, there are two particularly relevant separations.\cite{Reiher11,Reiher12,Reiher13,Reiher14,Reiher15,Ayers15a,Ayers15b,Bogus15,Szalay17,Reiher17b,Legeza18,Legeza19} To explain them, let us first observe that the underlying one-particle Hilbert space $\Ho$ consists of orbital and spin degrees of freedom, i.e.,
$\Ho \equiv \Ho_l \otimes \Ho_s$, where $\mbox{dim}(\Ho_l)\equiv d$, $\mbox{dim}(\Ho_s)\equiv 2$ and $\mbox{dim}(\Ho)\equiv D = 2 d$. The first partition picks one orbital $\ket{\chi}\in \Ho_l$ and then defines subsystem $A$ through the two spin-orbitals $\ket{\chi}\otimes \ket{\sigma}$, $\sigma = \uparrow,\downarrow$. Subsystem $B$ follows accordingly and comprises the remaining $D-2$ spin-orbitals. The corresponding measures for  entanglement and correlation can be referred to as single-orbital entanglement and correlation. As we will show in Section \ref{sec:formulae}, the fact that the total $N$-electron ground state of a molecular system is pure drastically simplifies the respective measures and in particular leads to closed formulas. The second more elaborated separation quantifies entanglement and correlation between two orbitals $\ket{\chi_i},\ket{\chi_j}\in \Ho_l$. This means to first trace out the complementary $D-4$ spin-orbitals to obtain a two-orbital reduced density matrix $\rho_{i,j}$ which is ``living'' on a sixteen-dimensional Fock space\cite{Reiher13,Bogus15} as illustrated in Figure \ref{fig:sectors}. Then, one applies to this new ``total state'' $\rho_{i,j}$ the formalism of bipartite entanglement and correlation for the separation $i \leftrightarrow j$ (see also the subsequent section).

Finally, let us also illustrate how the SSRs are implemented in the calculation of pairwise orbital entanglement. According to the previous comments, particularly Eq.~\eqref{eqn:tilde}, we just need to replace $\rho_{i,j}$ by its physical part $\rho_{i,j}^\textrm{Q}$. For the case of P-SSR and N-SSR this is illustrated in Figure \ref{fig:sectors}. $\rho_{i,j}^\textrm{P}$ is obtained by cutting out all light gray parts and $\rho_{i,j}^\textrm{N}$ follows after removing two additional entries (gray).

\section{Analytic Results}\label{sec:analytic}

\subsection{Closed Formulas for Entanglement and Correlation}\label{sec:formulae}
We first look at the single-orbital total correlation and entanglement and assume a pure quantum state $\rho =|\Psi\rangle\bra{\Psi}$ for the total $N$-electron system (typically it will be the ground state or an excited state of a molecular system). The one-orbital reduced density matrix associated with the orbital $\ket{\chi_j}$ is obtained by tracing out all remaining orbitals\cite{Sergey17}
\begin{equation}
\rho_j = \Tr_{\setminus \{j\}} [|\Psi\rangle\langle \Psi|] \label{eqn:one_orb_rdm}\,.
\end{equation}
We reiterate that the partial trace $\Tr_{\setminus \{j\}}[\cdot]$ does not mean to trace out particles but instead refers to the tensor product in the second quantization, i.e., it exploits the structure $\mathcal{F} = \mathcal{F}_j \otimes \mathcal{F}_{\setminus \{j\}}$.
From a practical point of view, the non-vanishing entries of the single-orbital reduced density matrix can be determined by calculating
expectation values of $\ket{\Psi}$ involving only fermionic creation and annihilation operators referring to orbital $\ket{\chi_j}$. For more details we refer the reader to Refs.~\onlinecite{Reiher13,Bogus15,Sergey17}. Due to the fixed particle number and the spin symmetry of $\ket{\Psi}$
the one-orbital reduced density matrix will be always diagonal in the local reference basis
$\{\ket{\Omega},\ket{\!\uparrow},\ket{\!\downarrow},\ket{\!\uparrow\downarrow}\}$
of orbital $\ket{\chi_j}$:
\begin{equation}
\rho_j = \begin{pmatrix}
p_1 & 0 & 0 & 0
\\
0 & p_2 & 0 & 0
\\
0 & 0 & p_3 & 0
\\
0 & 0 & 0 & p_4
\end{pmatrix}. \label{eqn:1rdm}
\end{equation}
By referring to the so-called Schmidt decomposition, the total state then takes the form
\begin{eqnarray}
|\Psi\rangle &=&\sqrt{ p_1} |\Omega\rangle \otimes |\Psi_{N,M}\rangle + \sqrt{p_2} |\!\uparrow\rangle \otimes |\Psi_{N-1,M-\frac{1}{2}}\rangle
\\
& &+ \sqrt{p_3} |\!\downarrow\rangle \otimes |\Psi_{N-1,M+\frac{1}{2}}\rangle + \sqrt{p_4} |\! \uparrow\downarrow\rangle \otimes |\Psi_{N-2,M}\rangle,\nonumber
\end{eqnarray}
where $|\Psi_{n,m}\rangle$ is a quantum state with particle number $n$ and magnetization $m$ of the complementary subsystem comprising the remaining $D-2$ spin-orbitals. Now we can readily determine the physical part $\rho^\textrm{Q}$ in the presence of P-SSR or N-SSR. In the absence of SSRs, the single-orbital entanglement of $\ket{\Psi}$ is simply given by the von Neumann entropy of $\rho_j$, and the single-orbital total correlation is simply twice the entanglement,
\begin{equation}
\begin{split}
E(|\Psi\rangle\langle\Psi|) &= S(\rho_j) = - \sum_{i=1}^4 p_i \ln(p_i). \label{eqn:single_noSSR}
\\
I(|\Psi\rangle\langle\Psi|) &= 2 E(|\Psi\rangle\langle\Psi|).
\end{split}
\end{equation}

In the case of Q-SSR, we need to consider the physical part $\rho^\text{Q}$ of $\rho = |\Psi\rangle\langle \Psi|$, which is no longer a pure state. Consequently the single-orbital entanglement cannot be quantified by the von Neumann entropy of $\rho_j$ anymore. Instead we have to invoke the geometric picture in Figure \ref{fig:states}. We first calculate the physical states with respect to P-SSR and N-SSR according to \eqref{eqn:tilde}, and then their correlation and entanglement are quantified using \eqref{eqn:mutual_info} and \eqref{eqn:rel_ent}. Explicit derivations are presented in Appendix \ref{app:single}. Remarkably, despite the fact that $\rho^{\textrm{Q}}$ is not a pure state anymore
the single-orbital correlation and entanglement under P-SSR and N-SSR still involves the spectrum of $\rho_j$ only:
\begin{equation}
\begin{split}
I(\rho^\textrm{P}) &= (p_1 + p_4) \ln(p_1 + p_4) + (p_2 + p_3)\ln(p_2+p_3)
\\
& - 2(p_1 \ln(p_1) + p_2 \ln(p_2) + p_3 \ln(p_3) + p_4 \ln(p_4)),
 \\
 I(\rho^\textrm{N}) & = p_1 \ln(p_1) + (p_2 + p_3)\ln(p_2+p_3) + p_4 \ln(p_4)
\\
 &- 2(p_1 \ln(p_1) + p_2 \ln(p_2) + p_3 \ln(p_3) + p_4 \ln(p_4)),
 \\
 E(\rho^\textrm{P}) &= (p_1 + p_4) \ln(p_1 + p_4) + (p_2 + p_3)\ln(p_2+p_3)
\\
& - p_1 \ln(p_1) - p_2 \ln(p_2) - p_3 \ln(p_3) - p_4 \ln(p_4),
\\
E(\rho^\textrm{N}) &= (p_2 + p_3) \ln(p_2 + p_3) - p_2 \ln(p_2) - p_3 \ln(p_3). \label{eqn:single_SSR}
\end{split}
\end{equation}
In particular, this implies immediately for both SSRs (Q=P,N)
\begin{equation}\label{eqn:IvsEpure}
I^\textrm{Q-SSR}(\rho) = E^\textrm{Q-SSR}(\rho) + E(\rho).
\end{equation}
For the case of no SSR, this is consistent with Eq.~\eqref{eqn:single_noSSR}.

As already explained in the previous section, the second particularly relevant partitioning of the total system leads to a notion of
orbital-orbital correlation and entanglement. It is fully described by
the two-orbital reduced density matrix associated with orbital $\ket{\chi_i}$ and $\ket{\chi_j}$
\begin{equation}
\rho_{i,j} = \Tr_{\setminus \{i,j\}} [|\Psi\rangle\langle\Psi|]. \label{eqn:two_orb_rdm}
\end{equation}
For the specific case of a total system consisting of just two orbitals, the only two-orbital ``reduced'' density operator is given by the total (pure) state. Consequently, the orbital-orbital correlation and entanglement thus coincide with the single-orbital ones and the above results \eqref{eqn:single_noSSR},\eqref{eqn:single_SSR} immediately apply.
Due to the electron interaction, the two-orbital reduced density matrices $\rho_{i,j}$ of \emph{general} systems are, however, not pure anymore.
This makes the calculation of orbital-orbital entanglement and classical correlation highly non-trivial: According to the definition of the relative entropy of entanglement \eqref{eqn:rel_ent} one needs to minimize the distance between $\rho_{i,j}$ and $\sigma_{i,j} \in \mathcal{D}_{sep}$ which \emph{a priori} involves 255 parameters. Yet, $\rho_{i,j}$ inherits particle and spin symmetries from the molecular ground state $\rho= \ket{\Psi}\bra{\Psi}$ which changes the general situation drastically. As a consequence almost all of its entries vanish as it is shown in Figure \ref{fig:sectors} (see also Refs.~\onlinecite{Reiher13}). Based on elaborated ideas the respective minimization \eqref{eqn:rel_ent} can be simplified accordingly by transferring those symmetries to $\mathcal{D}_{sep}.$\cite{vollbrecht2001entanglement,LexinQIT} The latter simplification eventually allows us to calculate in our work the entanglement between $\ket{\chi_1},\ket{\chi_2}\in \Ho_l$ for any $\rho_{i,j}$.


In the following, we will illustrate those concepts in the form of several analytical examples.

\subsection{Single Electron State}
At first glance it might seem somewhat bizarre to examine the correlation and entanglement in a system with only one particle. However, the reader shall bear in mind that the separation into subsystems is not referring to particles but \textit{orbitals}. The total Fock space $\mathcal{F}$ in our case of two orbitals ($\ket{1}, \ket{2}$) has indeed a natural tensor product structure between the Fock spaces of the first and second orbital, i.e., $\mathcal{F} = \mathcal{F}_1\otimes \mathcal{F}_2$. Therefore the notion of correlation and entanglement between two physically distinct orbitals is entirely legitimate even in the case of one single electron.

In the following we consider the specific one-electron state
\begin{equation}
|\Psi\rangle = \frac{1}{\sqrt{2}} (f^\dagger_{1\uparrow}  + f^\dagger_{2\uparrow})|{\Omega}\rangle \equiv\frac{|\!\uparrow\rangle \otimes |\Omega_2\rangle + |\Omega_1\rangle \otimes |\!\uparrow\rangle}{\sqrt{2}}. \label{eqn:one_electron}
\end{equation}
Here, $f^\dagger_{j\sigma}$ denotes the fermionic creation operator for the spin-orbital $\ket{j\sigma}$, $j=1,2, \sigma=\uparrow,\downarrow$ and $\ket{\Omega}$ and $\ket{\Omega_{1/2}}$ the global and local vacuum states, respectively.
If the SSRs are ignored, state \eqref{eqn:one_electron} is certainly entangled. Yet, as we will show now this entanglement is artificial since it disappears when the P-SSR is taken into account. To prove this, recall that the P-SSR physical part of $\rho$ is obtained by projecting onto fixed local parity sectors
\begin{equation}
\begin{split}
\rho^\textrm{P} &= \sum_{\tau, \tau' = \textrm{odd, even}} P_\tau \otimes P_{\tau'} \rho P_\tau \otimes P_{\tau'}
\\
&= \frac{1}{2} |\Omega_1\rangle\langle\Omega_1| \otimes |\!\uparrow\rangle\langle\uparrow\!| + \frac{1}{2} |\!\uparrow\rangle\langle\uparrow\!| \otimes  |\Omega_2\rangle\langle\Omega_2|, \label{eqn:rho2P}
\end{split}
\end{equation}
which is correlated but not entangled. Indeed, it is a classical mixture of two uncorrelated states.

For the sake of completeness, we would like to stress that for single electron states there is no difference between P-SSR and N-SSR. In particular for the state \eqref{eqn:one_electron} we find
\begin{eqnarray} \label{eqn:rho2N}
\rho^\textrm{N} &=& \sum_{m,n=0}^2 P_m \otimes P_n \,\rho\, P_m \otimes P_n \\
&=& \frac{1}{2} |\Omega_1\rangle\langle\Omega_1| \otimes |\!\uparrow\rangle\langle\uparrow\!| + \frac{1}{2} |\!\uparrow\rangle\langle\uparrow\!| \otimes  |\Omega_2\rangle\langle\Omega_2|= \rho^\textrm{P} . \nonumber
\end{eqnarray}

We present in Table \ref{tab:one_electron} the orbital total correlation (``Total''), entanglement (``Quantum'') and classical correlation (``Classical'') between $\ket{1}$ and $\ket{2}$ contained in state \eqref{eqn:one_electron} which can easily be calculated based on the physical states \eqref{eqn:rho2P}, \eqref{eqn:rho2N}.
\begin{table}[h]
\begin{tabular}{|c|c|c|c|}
\hline
\rule{0pt}{2.6ex}\rule[-1.2ex]{0pt}{0pt}
 & No SSR & P-SSR & N-SSR
\\
\hline
\rule{0pt}{2.6ex}\rule[-1.2ex]{0pt}{0pt}
Total & $2\ln(2)$ & $\ln(2)$ & $\ln(2)$
\\
\hline
\rule{0pt}{2.6ex}\rule[-1.2ex]{0pt}{0pt}
Quantum & $\ln(2)$ & 0 & 0
\\
\hline
\rule{0pt}{2.6ex}\rule[-1.2ex]{0pt}{0pt}
Classical & $0.208\ln(2)$ & $\ln(2)$ & $\ln(2)$
\\
\hline
\end{tabular}
\caption{Total, quantum and classical correlation between the two orbitals in the one-electron state $|\Psi\rangle$ in \eqref{eqn:one_electron}, for the case without SSR, with P-SSR and with N-SSR.}
\label{tab:one_electron}
\end{table}

The number $0.208$ in Table \ref{tab:one_electron} is the constant $\ln(4/3)/2$. When P-SSR or N-SSR is present (they are equivalent in the case of only one electron), all entanglement is wiped out and all correlation is classical, as it is shown by the second and third column. This striking example shows that one can never extract any entanglement from a single one-electron quantum state even if it appears at first sight as being entangled.

\subsection{Dissociated Hydrogen}\label{sec:diss_H2}
Having studied the orbital-orbital correlation and entanglement in a one-electron state, we now add a second electron to our two-orbital system. As an example, we consider the ground state of the hydrogen molecule in the dissociation limit. The two orbital system now consists of the 1s orbital at each nucleus (both orthonormal as we assume almost infinite separation) and the ground state follows as
\begin{equation}
|\Psi\rangle = \frac{1}{\sqrt{2}} (f^\dagger_{1\uparrow}f^\dagger_{2\downarrow}-f^\dagger_{1\downarrow}f^\dagger_{2\uparrow})|{\Omega}\rangle. \label{eqn:diss_H2}
\end{equation}
\begin{table}[h!]
\begin{tabular}{|c|c|c|c|}
\hline
\rule{0pt}{2.6ex}\rule[-1.2ex]{0pt}{0pt}
 & No SSR & P-SSR & N-SSR
\\
\hline
\rule{0pt}{2.6ex}\rule[-1.2ex]{0pt}{0pt}
Total & $2\ln(2)$ & $2\ln(2)$  & $2\ln(2)$
\\
\hline
\rule{0pt}{2.6ex}\rule[-1.2ex]{0pt}{0pt}
Quantum & $\ln(2)$ & $\ln(2)$ & $\ln(2)$
\\
\hline
\rule{0pt}{2.6ex}\rule[-1.2ex]{0pt}{0pt}
Classical & $0.208\ln(2)$ & $0.208\ln(2)$ & $0.208\ln(2)$
\\
\hline
\end{tabular}
\caption{Total, quantum and classical correlation between both orbitals $\ket{1},\ket{2}$ in the dissociated hydrogen ground state $|\Psi\rangle$ in \eqref{eqn:diss_H2}, for the case without SSR, with P-SSR and N-SSR.}
\label{tab:diss_H2}
\end{table}

\noindent In Table \ref{tab:diss_H2} we list the total correlation, entanglement and classical correlation between $\ket{1}$ and $\ket{2}$. When no SSR is considered, all three types of correlation equal those of the one-electron state in Table \ref{tab:one_electron}. However, in contrast to the latter, the ground state $|\Psi\rangle$ of the dissociated hydrogen molecule is already a physical state, with respect to both P-SSR and N-SSR. From \eqref{eqn:diss_H2} we infer that $|\Psi\rangle$ is a pure state with definite local parities $(\textrm{odd},\textrm{odd})$ and definite local particle numbers $(1,1)$. The projection \eqref{eqn:tilde} therefore does not change the state $|\Psi\rangle\bra{\Psi}$ and thus all three types of correlation are unaffected by P-SSR and N-SSR according to \eqref{eqn:SSRmeasures}.

\subsection{Hubbard Dimer}\label{sec:Hubbard}
The dissociated hydrogen molecule described in the previous section is a very special case with a definite local particle number (and of course, parity). If we consider intermediate bond length, however, different local particle number or parity sectors will start to mix, and hence the behavior of the orbital correlation and entanglement will be more interesting. To elaborate on this, we turn to an elementary model system.  The Hamiltonian of this Hubbard dimer which comprises two sites reads
\begin{equation}\label{eqn:HHubb}
H = - t \sum_{\sigma = \uparrow, \downarrow} f^\dagger_{1\sigma}f^{\phantom{\dagger}}_{2\sigma} + \textrm{H.c.} + U\sum_{j=1,2}\hat{n}_{j\uparrow}\hat{n}_{j\downarrow},
\end{equation}
where $f^{(\dagger)}_{j\sigma}$ are annihilation (creation) operators associated with a spin $\sigma$ electron on site $j=1,2$. It is worth recalling that small hopping rates $t$ correspond to larger inter-nuclei separations. The repulsive potential $U$ penalizes any doubly occupied site, effectively describing the Coulomb repulsion of two electrons in the same 1s orbital. Exploiting the symmetries of \eqref{eqn:HHubb} one easily determines the ground state of the Hubbard dimer:
\begin{equation}
\begin{split}
|\Psi\rangle &= \frac{a}{\sqrt{2}}(f^\dagger_{1\uparrow}f^\dagger_{2\downarrow}-f^\dagger_{1\downarrow}f^\dagger_{2\uparrow})|\Omega\rangle \\\
& \quad + \frac{b}{\sqrt{2}}(f^\dagger_{1\uparrow}f^\dagger_{1\downarrow}-f^\dagger_{2\downarrow}f^\dagger_{2\uparrow})|\Omega\rangle , \label{eqn:hubbard_gs}
\end{split}
\end{equation}
where
\begin{equation}
\begin{split}
a &= \sqrt{\frac{W + \frac{U}{2}}{2W}}, \quad b= \frac{2t}{\sqrt{2W\left(W+\frac{U}{2}\right)}},
\\
W &= \sqrt{\frac{U^2}{4}+4t^2}.
\end{split}
\end{equation}

\begin{figure}[h]
\centering
\includegraphics[scale=0.27]{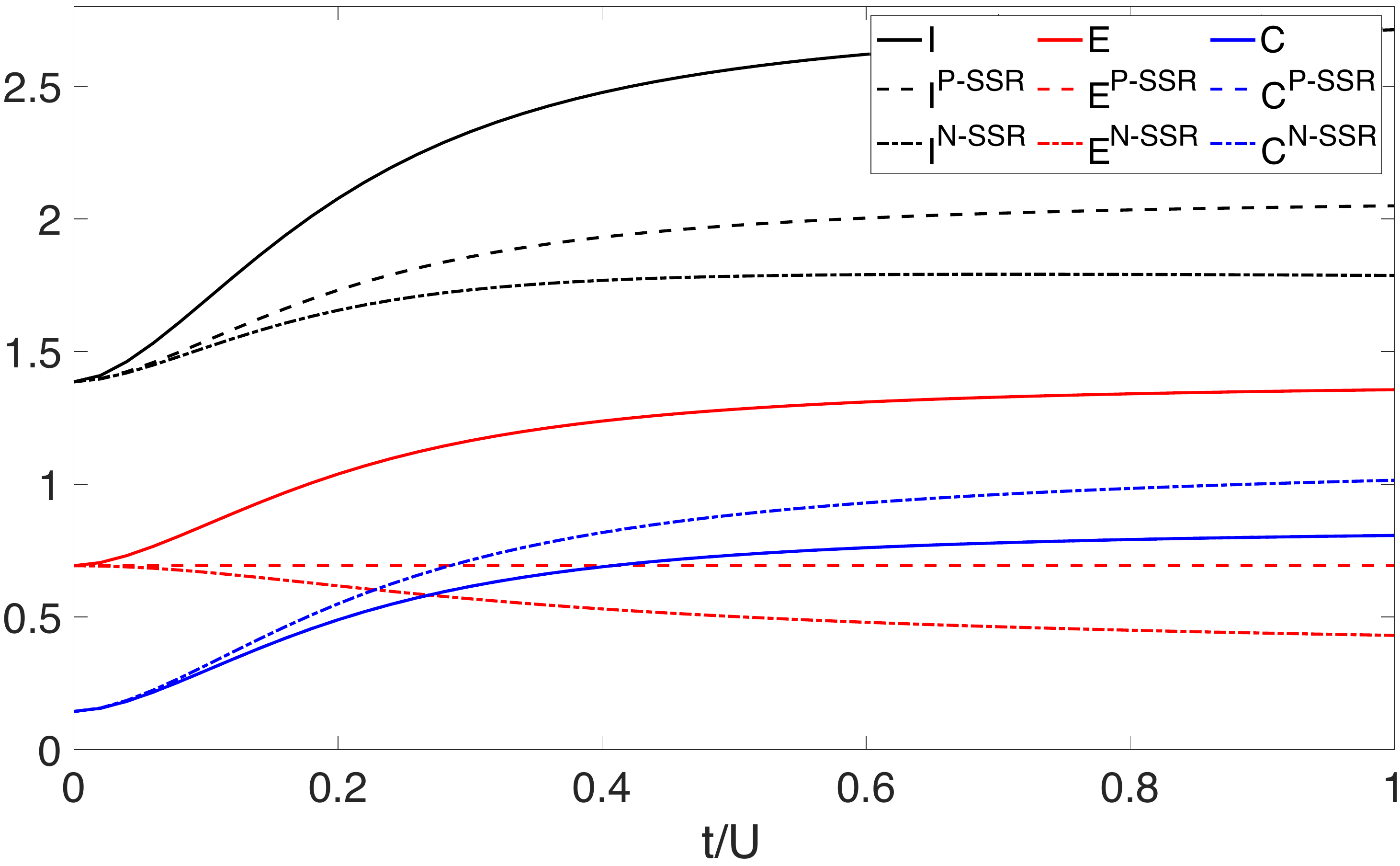}
\caption{Total correlation $I$ (black), entanglement $E$ (red) and classical correlation $C$ (blue) between both sites for the ground state of the Hubbard dimer \eqref{eqn:HHubb} as functions of the ratio $t/U$. The curves for $C$ and $C^{\textrm{P-SSR}}$ coincide.}
\label{fig:dimer}
\end{figure}

The orbital-orbital total correlation $I$, entanglement $E$ and classical correlation $C$ in the ground state \eqref{eqn:hubbard_gs} are plotted in Figure \ref{fig:dimer}, for the case without SSR, with P-SSR and N-SSR\cite{Lexin20a}, as a function of the parameter $t/U$. In this special case of just two orbitals in total and a pure state, the single-orbital and orbital-orbital correlation and entanglement coincide.

In the case without any SSR, the orbital-orbital entanglement $E$ (red solid) is exactly half of the total correlation $I$. For finite $t/U$, i.e., when $b > 0$ in \eqref{eqn:hubbard_gs}, P-SSR and N-SSR drastically reduce the orbital-orbital total correlation and entanglement, as is shown by the curves corresponding to $I^\textrm{P-SSR}$ (black dashed), $I^\textrm{N-SSR}$ (black dot-dashed), $E^\textrm{P-SSR}$ (red dashed) and $E^\textrm{N-SSR}$ (red dot-dashed). N-SSR being the stronger rule, reduces correlation and entanglement the most. However, when we take $t/U \rightarrow 0$, corresponding to the dissociation limit, the effect of P-SSR and N-SSR disappears. This is due to the vanishing coefficient $b$ in \eqref{eqn:hubbard_gs} in the dissociation limit, which results in a ground state that is physical in the presence of both P-SSR and N-SSR, as it has already been pointed out in Section \ref{sec:diss_H2}.

All these previous elementary examples already reveal the drastic effect of SSRs on orbital correlation and entanglement.
In the following section we will apply the quantum information theoretical concepts to systems with more orbitals.

\section{Numerical results}\label{sec:results}
In this section we investigate the correlation and entanglement in the ground states of molecules. We consider exemplarily three chemical systems, the water molecule $\mathrm{H_2O}$, naphthalene $\mathrm{C_{10}H_8}$ and the chromium dimer $\mathrm{Cr_2}$, each containing different levels of correlation. Accurate ground states are found by using the density matrix renormalization group (DMRG) method as outlined in the following Section \ref{sec:comp}. The single-orbital and orbital-orbital correlation and entanglement are studied in Sections \ref{sec:results1} and \ref{sec:results2}. To avoid any confusion, it is worth noticing that the choice of orbitals with respect to which those quantities are eventually calculated is made only \emph{after} having obtained a good approximation of the molecule's quantum state. Only for illustrative purposes we will choose in the following for this the Hartree-Fock orbitals which at the same time already play some role in the calculation of the ground state. Hence, in complete analogy to expectation values of more conventional observables, the orbital entanglement and correlation depend on both the molecule's quantum state $\ket{\Psi}$ and the choice of orbitals. Yet, they do not depend on the numerical method that is used to obtain the concrete quantum state $\ket{\Psi}$.

\subsection{Computational details} \label{sec:comp}
In order to find a ground state, and from it compute the required orbital reduced density matrices, we start with a preceding Hartree-Fock calculation to establish the molecular orbitals. For our post-Hartree-Fock DMRG calculation we construct an active space consisting of the most relevant molecular orbitals and compute integral elements with the one- and two-particle Hamiltonian $T$ and $V$, respectively. Those respective integral elements define the electronic Hamiltonian at hand by referring to second quantization:
\begin{align}\label{eqn:ham}
    H &= \sum\limits_{ij\sigma}
    T_{ij} c^\dagger_{i\sigma} c_{j\sigma}^{\phantom{\dagger}} + \sum\limits_{ijkl\sigma \tau} V_{ijkl} c^\dagger_{i\sigma} c^\dagger_{j\tau} c_{k\tau}^{\phantom{\dagger}} c_{l \sigma}^{\phantom{\dagger}}\,.
\end{align}
For our DMRG calculations we do not fix any molecular symmetries. Yet, the total particle number and the z-component of the total spin are always assumed to be conserved, with the latter one always fixed to be zero. Furthermore, we did not employ exceedingly large active spaces for two reasons. First, the ground states are almost exactly found, as it is shown below. Second, for the purpose of demonstration, our findings do not qualitatively hinge on the tiny improvement found by resorting to larger active spaces. In particular, the reduction in correlation and entanglement due to the regularly ignored superselection rules turns out to exceeds by several orders of magnitude the truncation error of our active spaces.

To obtain for each ground state $\ket{\Psi}$ the required one- ($\rho_j$) and two-orbital reduced density matrices ($\rho_{i,j}$) we trace all orbitals except one and two, respectively (recall Eqs.~\eqref{eqn:one_orb_rdm},\eqref{eqn:two_orb_rdm}).
In addition, we determine the one-particle reduced matrix
\begin{align}
   \gamma_{i\sigma, j\tau} \equiv \bra{\Psi} c^\dagger_{j\tau} c^{\phantom{\dagger}}_{i\sigma} \ket{\Psi}, \quad \sigma,\tau = \uparrow, \downarrow, \label{eqn:one_part}
\end{align}
whose eigenvalues (\emph{natural occupation numbers}) provide insights into the reference basis-independent intrinsic correlation (as we will discuss below).
Notice that $\gamma$ is trace-normalized to the particle number $N$.

\begin{figure}[!t]
    \centering
    \includegraphics[scale=0.29]{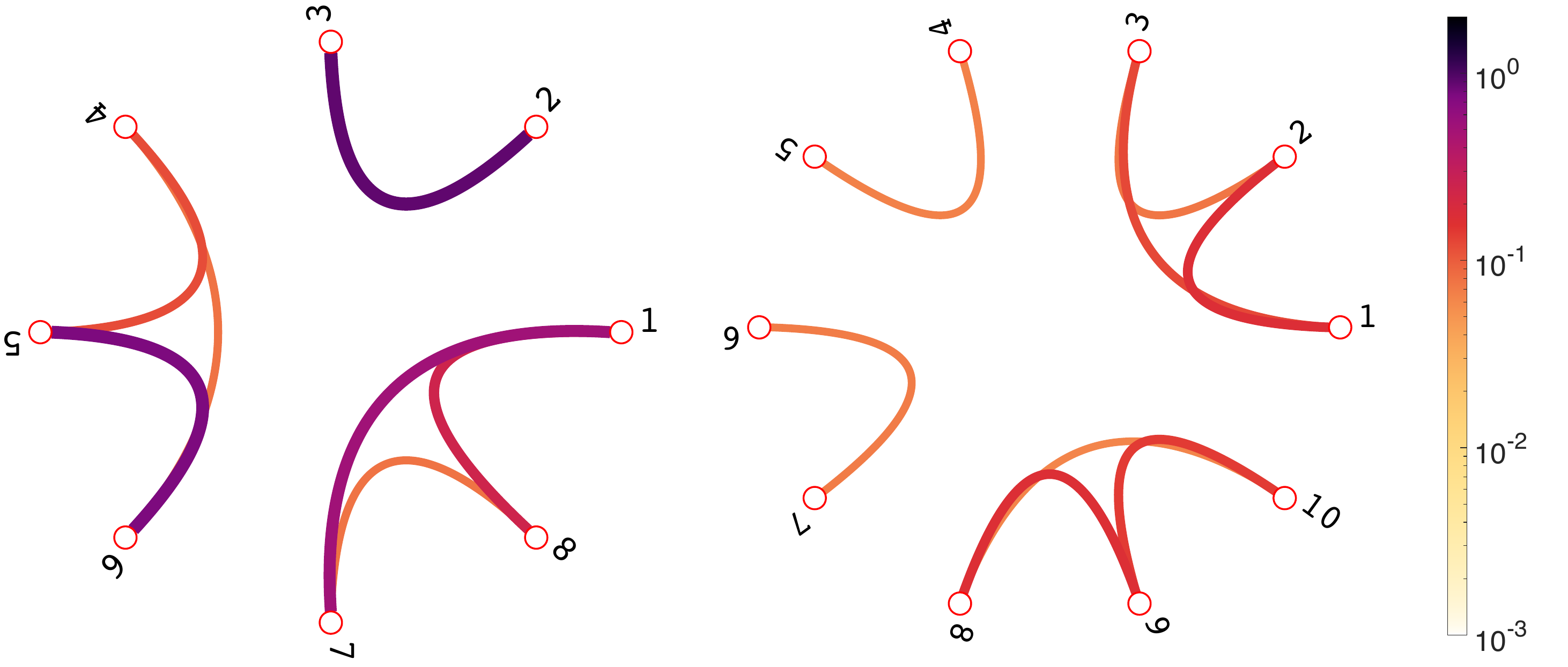}
    \caption{Matrix elements $|T_{ij}|$ of the one-particle Hamiltonian with respect to the Hartree-Fock orbitals for $\mathrm{H_2O}$ (left) and $\mathrm{C_{10}H_8}$ (right).}
    \label{fig:OneHam}
\end{figure}

The computations for the water molecule were done using $(6e,8o)$, meaning a total of six active electrons and eight active orbitals with a $6-31G(d)$ basis set. Other orbitals are classified as either frozen (doubly occupied) or virtual (hosting zero electrons). The inter-atomic distance between the oxygen and the hydrogen atoms is given by $\SI{0.7567}{\angstrom}$ while the angle between the atoms is $\ang{118}$. The system has converged fully to its FCI energy $E/\si{\hartree}=-75.78608737$ and therefore the DMRG wavefunction can be considered as being exact. The system has a $C_{2v}$ symmetry, i.e.,  it is symmetrical with respect to rotations around the oxygen atom in a plane and also has a reflection symmetry with respect to the horizontal and vertical planes. This decomposes the orbitals into subgroups labeled by irreducible representations of this point group, whose one particle Hamiltonian matrix elements do not couple to each other. The active Hartree-Fock orbitals $\{1,7,8\}$ belong to the $B_1$ representation, while $\{2,3\}$ belong to the $A_2$ representation, and the remaining orbitals $\{4,5,6\}$ belong to the $B_2$ representation. We remark that the Hartree-Fock orbitals were reordered in order to showcase these irreducible representations, as is illustrated in Figure \ref{fig:OneHam} (left). A similar reordering is also performed for the naphthalene molecule.

\begin{figure*}[htb]
\centering
\includegraphics[scale=0.405]{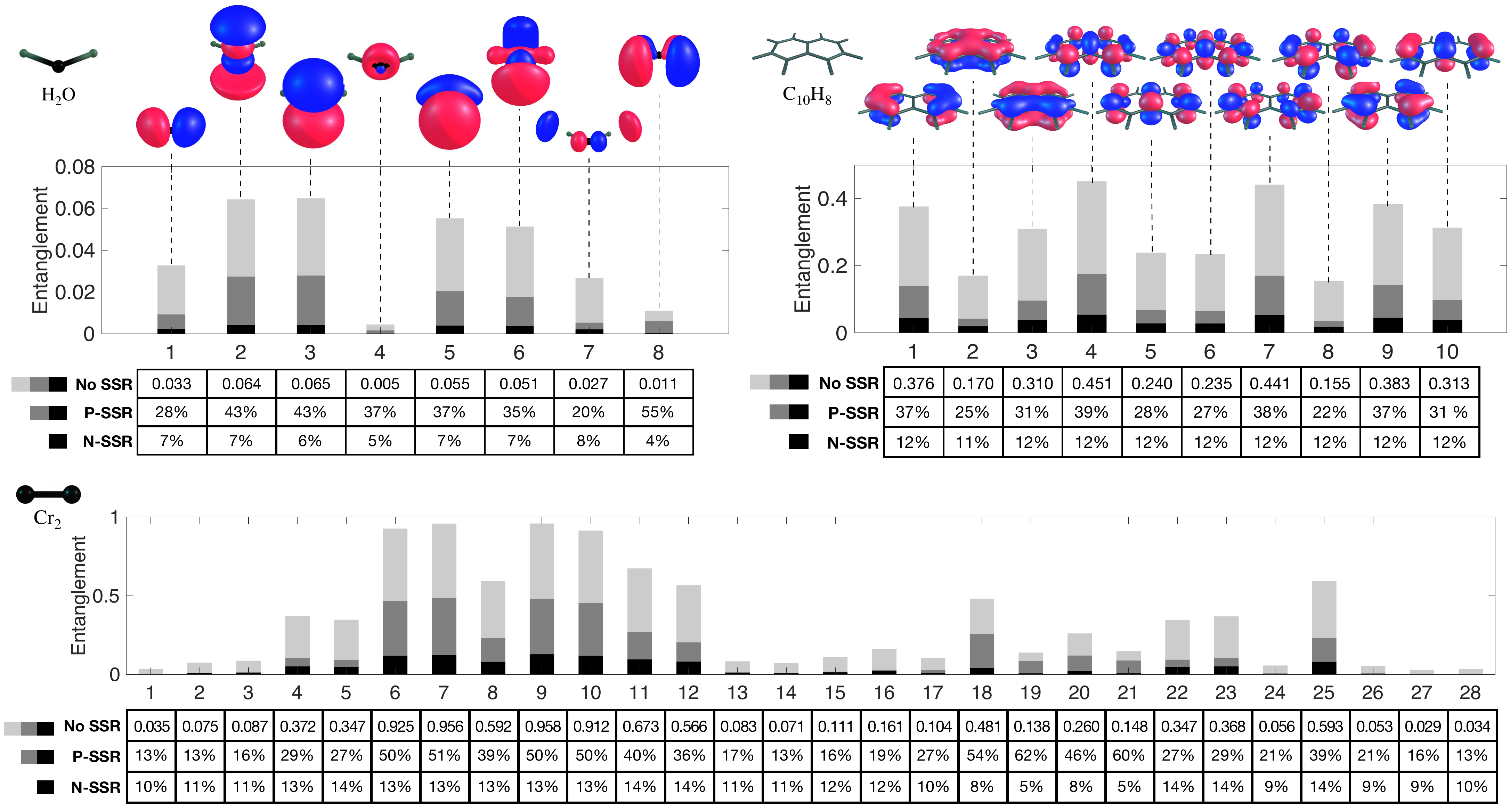}
\caption{Single-orbital entanglement of the Hartree-Fock orbitals (as visualized) in the ground states of $\mathrm{H_2O}$, $\mathrm{C_{10}H_8}$ and $\mathrm{Cr_2}$ for the three cases of no, P- and N-SSR. Exact values of entanglement and the remaining entanglement in terms of percentage of the No SSR case in the presence of P-SSR and N-SSR are listed in the table below each plot.}
\label{fig:1rdm}
\end{figure*}

The naphthalene molecule was done in a $(10e,10o)$ set-up with a $cc-pVDZ$ basis. The inter-atomic distances vary depending on the bond between $\SI{1.3701}{\angstrom}$ and $\SI{1.4277}{\angstrom}$ and angles between $\ang{118}$ and $\ang{120}$. With a final energy of $E/\si{\hartree}=-383.50563350$ it might also be regarded as converged. The system has a $D_{2h}$ symmetry meaning there is a rotation and inversion symmetry and also invariance under reflection. Here the orbitals $\{1,2,3\}$ belonging to $B_{3u}$, $\{4,5\}$ belonging to $B_{2g}$, $\{6,7\}$ belonging to $A_u$ and $\{8,9,10\}$ belonging to $B_{1g}$, split the system into irreducible representations, as is shown in Figure \ref{fig:OneHam} (right).

Finally, for the chromium dimer with a separation distance of $\SI{1.68}{\angstrom}$ a much larger active space of $(12e,28o)$ depicted in a ANO-RCC-VTZP+RX2C basis is used due to recover its increased correlation. The final DMRG energy based on a bond dimension $m=2048$ of the underlying matrix product state ansatz is $E/\si{\hartree} = -2099.30351824$. The symmetry group of this molecule is $D_{\infty h}$, i.e., the symmetry is much higher than for the other two molecules. Here the first six orbitals belong to $A_g$, orbitals $\{7,8,9\}$ to $B_{3u}$, $\{10,11,12\}$ to $B_{2u}$, $\{13,14\}$ to $B_{1g}$, $\{15,16,17,18,19,20\}$ to $B_{1u}$, $\{20, 21, 22, 23\}$ to $B_{2g}$, $\{24,25,26\}$ to $B_{3g}$ and $\{27,28\}$ to $A_u$.

A comment is in order concerning the accuracy of our numerical approach. It is true that our approach based on not particularly large active spaces does not allow us to recover most of the dynamic correlations. Yet, the by far most significant contribution to single and two-orbital entanglement and correlation will come from the static correlation. This is very similar to the fact that the bonding order of molecules can be determined by approaches which do not recover dynamic correlations (typically even much smaller active spaces than in our work would be sufficient for this). Moreover, the superselection rule which has erroneously been ignored in previous studies will have a much more significant impact on our results than the neglected part of the dynamic correlations. It is also crucial to notice that the bond dimensions of our DMRG calculations would in principle allow for very strong correlations. Just to illustrate this, even the tiny bond dimension of 2 would already allow one to describe one bond with maximal correlation (value $\approx 2.8$) between one pair of orbitals. As a second remark concerning the numerical quality, we would like to reiterate that the preceding Hartree-Fock calculations involved much larger basis sets involving several hundreds atomic orbitals. Only in the subsequent QC-DRMG calculations we restricted to smaller active space by carefully choosing \emph{molecular} orbitals (taking into account point group symmetries and the energies of the Hartree-Fock orbitals). We have also systematically varied the bond dimensions in QC-DMRG and the sizes of the underlying active spaces to confirm that the orbital entanglement and correlation patterns do not change anymore on the relevant scales.

\subsection{Single-orbital entanglement and correlation}\label{sec:results1}
After having obtained the ground states of the desired molecules, we can now explore single-orbital correlation and entanglement by applying the respective formulas from Section \ref{sec:formulae}. Since the states $\rho$ at hand are all \emph{pure} states, the single-orbital total correlation without any SSR is always exactly twice the single-orbital entanglement, as stated in \eqref{eqn:single_noSSR}. When P-SSR or N-SSR is taken into account, the respective physical states $\rho^\textrm{P}$ and $\rho^\textrm{N}$ are no longer pure, but in general mixtures of fixed parity or particle number states. However, in the form of Eq.~\eqref{eqn:IvsEpure} there still exists an exact relation between total correlation and entanglement in the presence of SSRs. Because of this, we focus in this section on the entanglement.

In Figure \ref{fig:1rdm} we plotted the single-orbital entanglement in the ground state of the $\mathrm{H_2O}$, $\mathrm{C_{10}H_8}$ and $\mathrm{Cr_2}$, respectively, for the case without SSR, with P-SSR and with N-SSR, using the analytic formulas \eqref{eqn:single_noSSR} and \eqref{eqn:single_SSR}. Below each figure we listed the exact values of single-orbital entanglement in the absence of SSRs, and also the remaining entanglement in the presence of P-SSR and N-SSR, in percentage.
All these results refer here and in the following to the Hartree-Fock orbitals which are for the sake of completeness also visualized for $\mathrm{H_2O}$ and $\mathrm{C_{10}H_8}$.

Generally speaking, the single-orbital entanglement of Hartree-Fock orbitals is quite small compared to the one of \emph{atomic} orbitals in a bond (see Sections \ref{sec:diss_H2} and \ref{sec:Hubbard}), particularly for $\mathrm{H_2O}$ and $\mathrm{C_{10}H_8}$. This confirms that the Hartree-Fock orbitals give rise to a much more local structure than that the atomic orbitals and in that sense define a much better starting point for high precision ground state methods.  Comparing the three systems, the water molecule contains the weakest single-orbital entanglement, less than $10^{-1}$ for all eight orbitals, whereas the strongest single-orbital entanglement in naphthalene and the chromium dimer have the values $0.451$ and $0.958$, respectively. This already emphasizes the different levels of correlation in those systems. Yet, it is worth noticing that any type of orbital entanglement and correlation (e.g., single- or two-orbital entanglement) strongly depends on the chosen reference basis. Even for a configuration state \eqref{config} one could find large orbital entanglement and correlation if one referred to orbitals which differ a lot from the natural orbitals.

To dwell a bit more on the concept of correlation, we emphasize that a basis set-independent notion can be defined in terms of the one-particle reduced density matrix $\gamma$ \eqref{eqn:one_part}. To explain this, we first observe that for configuration states \eqref{config} $\gamma$ has eigenvalues (\emph{natural occupation numbers}) all identical to one and zero, respectively. Arranging them in decreasing order gives rise to the so-called ``Hartree-Fock point'' $\vec{\lambda}^{\textrm{HF}} \equiv (1,\ldots,1,0,\ldots,0)$, where the first $N$ entries are $1$, and the remaining $D-N$ are $0$. The distance of the decreasingly ordered natural occupation numbers $\vec{\lambda}\equiv (\lambda_\alpha)_{\alpha=1}^D\equiv \mbox{spec}^\downarrow(\gamma)$ to $\vec{\lambda}^{\mathrm{HF}}$,
\begin{eqnarray}\label{eqn:lambdadist}
\mbox{dist}_1(\vec{\lambda},\vec{\lambda}^{\textrm{HF}})  &\equiv& \sum_{\alpha=1}^D |\lambda_\alpha - \lambda^{\textrm{HF}}_\alpha| \nonumber \\
&=& \sum_{\alpha=1}^N \left(1-\lambda_\alpha\right) + \sum_{\alpha=N+1}^D \lambda_\alpha\,,
\end{eqnarray}
defines an elementary measure for the \emph{intrinsic} (i.e., reference basis-independent) correlation of a quantum state $\rho$.
As a matter of fact, one easily proves\cite{CSthesis,CScorrmeas} that the overlap of an $N$-electron pure states $\ket{\Psi}$ with a configuration state built up from its $N$ first natural spin-orbitals $\ket{\varphi_\alpha}$ fulfills,
\begin{equation}
\frac{1}{2N} \mbox{dist}_1(\vec{\lambda},\vec{\lambda}^{\textrm{HF}}) \leq 1-\big|\!\langle \varphi_1,\ldots,\varphi_N\ket{\Psi}\!\big|^2 \leq \frac{1}{2} \mbox{dist}_1(\vec{\lambda},\vec{\lambda}^{\textrm{HF}})\,.
\end{equation}
This means that the maximized overlap of $\ket{\Psi}$ with a configuration state (Slater determinant) approaches the value one whenever $\vec{\lambda}$ is close to the ``Hartree-Fock point''. The closer a ground state $\ket{\Psi}$ is to the closest configuration state, the more accurate will be the Hartree-Fock approximation for the respective molecular system. Applying the measure \eqref{eqn:lambdadist} of intrinsic correlation to the ground states of water, naphthalene and chromium dimer yields the values $0.004$, $0.025$ and $0.084$, respectively. Those results comprehensively confirm that the systems at hand are not that strongly correlated and water in particular is weakly correlated.
As our analysis (partly inspired by Ref.~\onlinecite{Reiher11}) in the following section will show, the pairwise orbital entanglement and correlation pattern will be dominated by the point group symmetries of the one-particle Hamiltonian $T$ of the molecule \emph{as long as} the intrinsic correlation of the ground state is small enough.

From a quantum information perspective, the effect of SSRs on the single-orbital entanglement is drastic. The presence of P-SSR and N-SSR considerably reduces the amount of physical entanglement. According to the accompanying tables in Figure \ref{fig:1rdm}, P-SSR eliminates at least $45\%$ of it and occasionally even up to $87\%$. Taking into account the more relevant N-SSR eliminates between $86\%$ and $96\%$. Intriguingly, the entanglement hierarchy, however, remains almost intact. That is, if one orbital is more entangled with the rest than another orbital, the same will likely hold in the presence of P-SSR and N-SSR. It is also worth noting that even the stronger N-SSR does never wipe out the entire entanglement, which we shall see below can happen in the context of orbital-orbital entanglement.

From a quantum chemistry point of view, in Figure \ref{fig:1rdm} the single-orbital entanglement varies significantly from orbital to orbital. In particular, some orbitals are barely correlated with the others. Since we have chosen our active spaces systematically by taking into account various Hartree-Fock orbitals energetically closest to the Fermi level, this is a good indicator that our actives spaces were large enough to cover most of the correlation contained in the three molecules. On the other hand, if most orbitals were strongly entangled, the respective active space probably would have been too small. This is also the reason why the single-orbital correlation could help to automate the selection of active orbital spaces in quantum chemistry, as has been suggested and worked out in Refs.~\onlinecite{Reiher11,Reiher12}. Our refined analytic results \eqref{eqn:single_noSSR} and \eqref{eqn:single_SSR} demonstrated in Figure \ref{fig:1rdm} are able to identify exactly the quantum part of the total correlation while also taking into account the important superselection rules. These additional facets make precise the usage of quantum information theoretic concepts in the context of quantum chemistry, and may offer new perspectives into the selection of active space.

\subsection{Orbital-orbital entanglement and correlation}\label{sec:results2}
\begin{figure}[!t]
    \centering
    \includegraphics[scale=0.46]{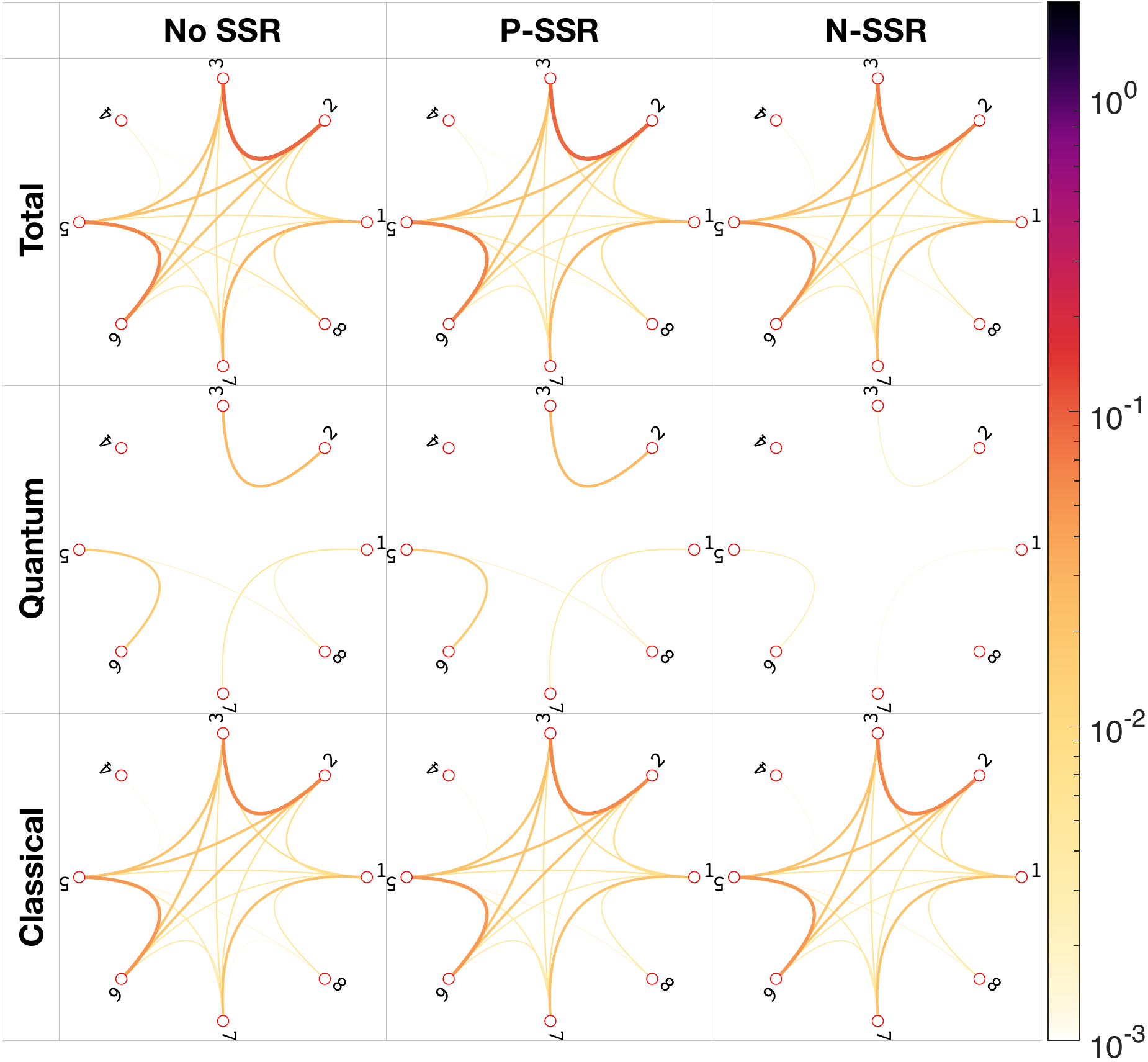}
    \caption{Total correlation, entanglement (``Quantum'') and classical correlation between any two Hartree-Fock orbitals in the ground state of $\mathrm{H_2O}$ for the case with no, P- and N-SSR.}
    \label{fig:CorrWater}
\end{figure}

To provide more detailed insights into the correlation and entanglement structure of molecular ground states, we also study the pairwise correlation and entanglement between two orbitals. This can be done in general in three steps: 1. Obtain the two-orbital reduced density matrix $\rho_{i,j}$ by tracing out all orbital degrees of freedom but orbital $i$ and $j$ as described in \eqref{eqn:two_orb_rdm}. 2. Apply the suitable projection to obtain the physical part $\rho_{i,j}^\textrm{Q}$ of $\rho_{i,j}$ under Q-SSR, as explained in Section \ref{sec:SSRincorp}. 3. Calculate the correlation and entanglement between the two orbital using \eqref{eqn:SSRmeasures}.

It is worth recalling that the two-orbital reduced density matrices $\rho_{i,j}$ are typically highly mixed, which is due to the coupling between different Hartree-Fock orbitals in the Hamiltonian \eqref{eqn:ham}. The total correlation for a mixed state, measured by the distance to the closest uncorrelated state \eqref{eqn:mutImin}, can always be calculated analytically, as it coincides with the quantum mutual information \eqref{eqn:mutual_info}. However, the entanglement quantified as the distance to the closest separable state \eqref{eqn:rel_ent} is immensely difficult to obtain by analytic means due to two reasons. One is the challenge imposed by the high dimensionality, even if we are interested in the entanglement between just two orbitals. The respective total system in that case has a Hilbert space isomorphic to $\mathbb{C}^4 \otimes \mathbb{C}^4$ (see also Figure \ref{fig:sectors}). A generic density matrix is then described by $16\times16-1 = 255$ real-valued parameters. In order to find the closest separable state to a two-orbital state $\rho_{i,j}$, one already needs to navigate through $255$ parameters. The second difficulty lies in the complexity of the boundary of the set of separable states $\mathcal{D}_{sep}$. Determining whether a state is separable or not is an NP-hard problem.\cite{gurvits2003classical} This also explains why exact criteria for separability are know so far only for Hilbert spaces with dimensions up to $2\times3$.\cite{horodecki1996separability,peres1996separability} In some cases when the total state exhibits many symmetries, the closest separable state for the two-orbital reduced state can still be found analytically.\cite{LexinQIT} In general, however, one has no choice but to resort to a combination of analytic tools and numerical methods, which is exactly what we did to obtain the results for orbital-orbital correlation and entanglement.

\begin{figure}[!t]
    \centering
    \includegraphics[scale=0.46]{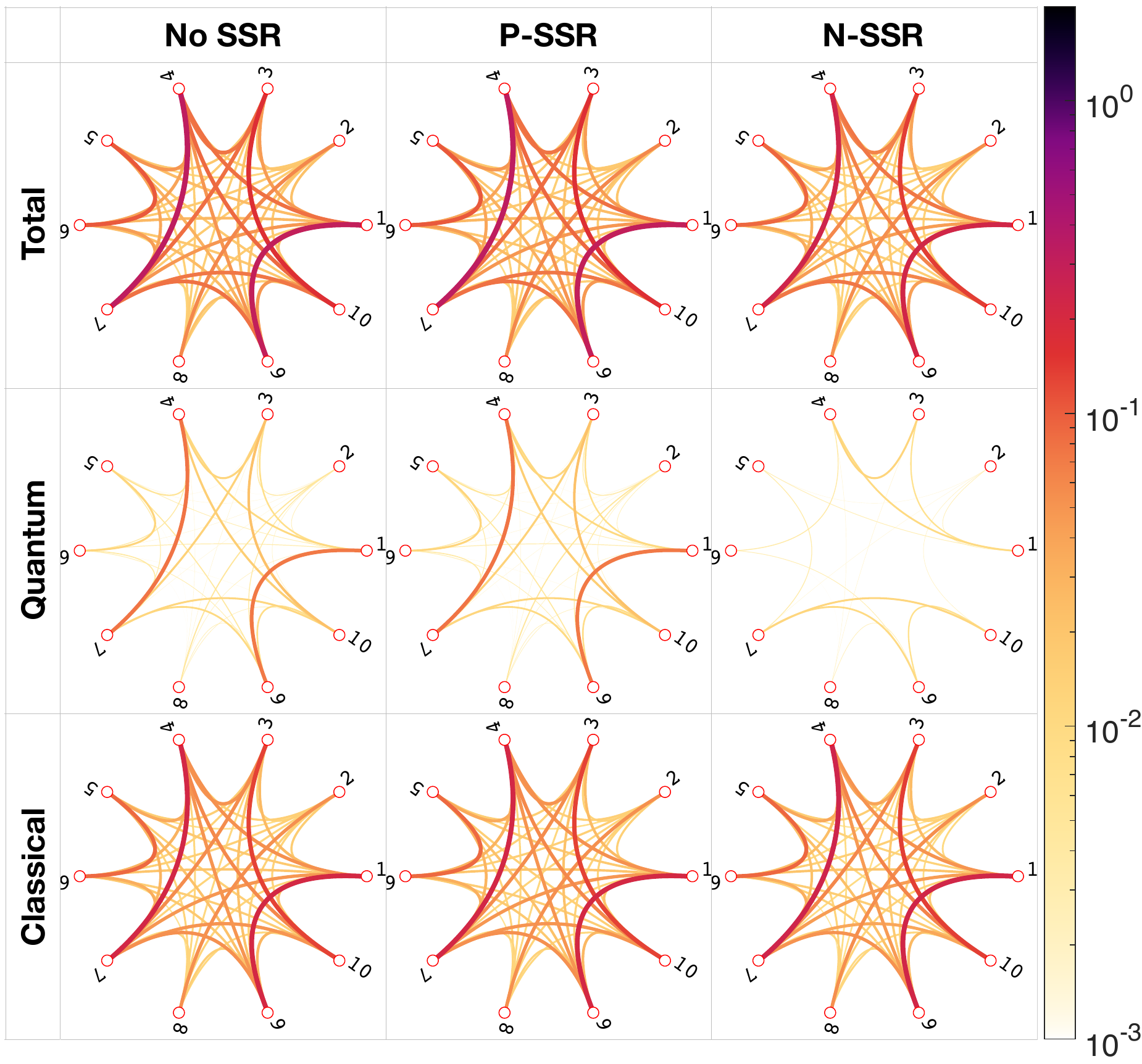}
    \caption{Total correlation, entanglement (``Quantum'') and classical correlation between any two Hartree-Fock orbitals in the ground state of $\mathrm{C_{10}H_8}$ for the case with no, P- and N-SSR.}
    \label{fig:CorrNaph}
\end{figure}

\begin{figure}[t]
    \centering
    \includegraphics[scale=0.46]{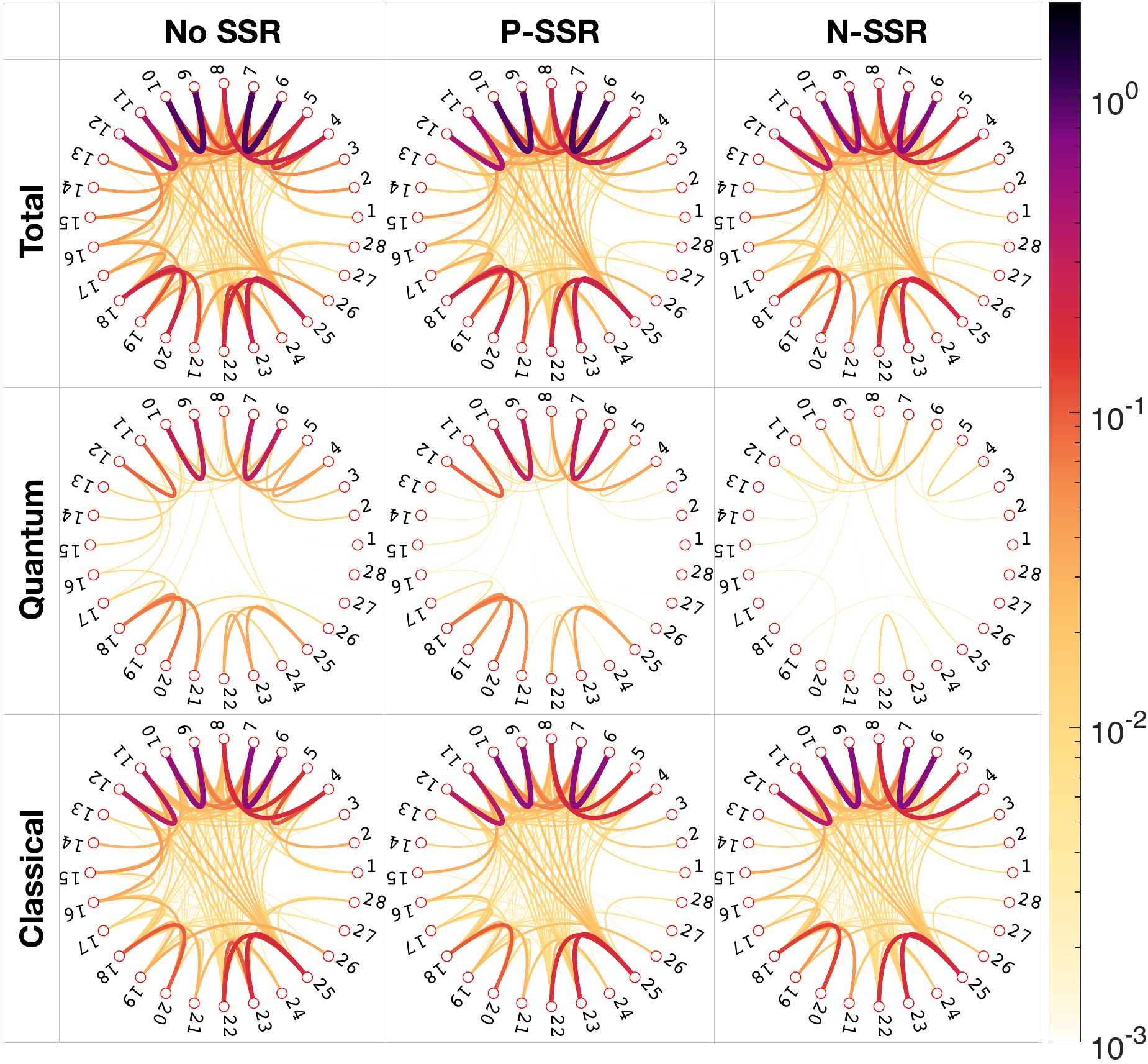}
    \caption{Total correlation, entanglement (``Quantum'') and classical correlation between any two Hartree-Fock orbitals in the ground state of $\mathrm{Cr_2}$ for the case with no, P- and N-SSR.}
    \label{fig:CorrCr2_28}
\end{figure}

The quantities we calculate are the total correlation, entanglement and classical correlation between two Hartree-Fock orbitals, for the case without SSR, with P-SSR and with N-SSR. All those nine quantities are calculate for all pairwise combinations of orbitals, for the ground states of all three molecules introduced in Section \ref{sec:comp}. Since each ground state is a singlet with a fixed electron number, any two-orbital reduced state $\rho_{i,j}$ is also symmetric with respect to the total two-orbital spin, magnetization and particle number.\cite{LexinQIT} Using the symmetry argument,\cite{vollbrecht2001entanglement} the closest separable state $\sigma^\ast_{i,j}$ is block diagonal in the simultaneous eigenbasis of the respective two-orbital spin and particle number operators (as also illustrated in Figure \ref{fig:sectors}). In the case of N-SSR, the projections used for calculating the physical state further increase the symmetry of $\sigma^\ast_{i,j}$, which eventually allows us to determine it analytically.\cite{LexinQIT} For the case without SSR and with P-SSR, we developed an algorithm based on semidefinite programming to find the closest separable state and calculate the entanglement in an numerically exact way.\cite{LexinQIT}

\begin{figure*}[htb]
\centering
\includegraphics[scale=0.45]{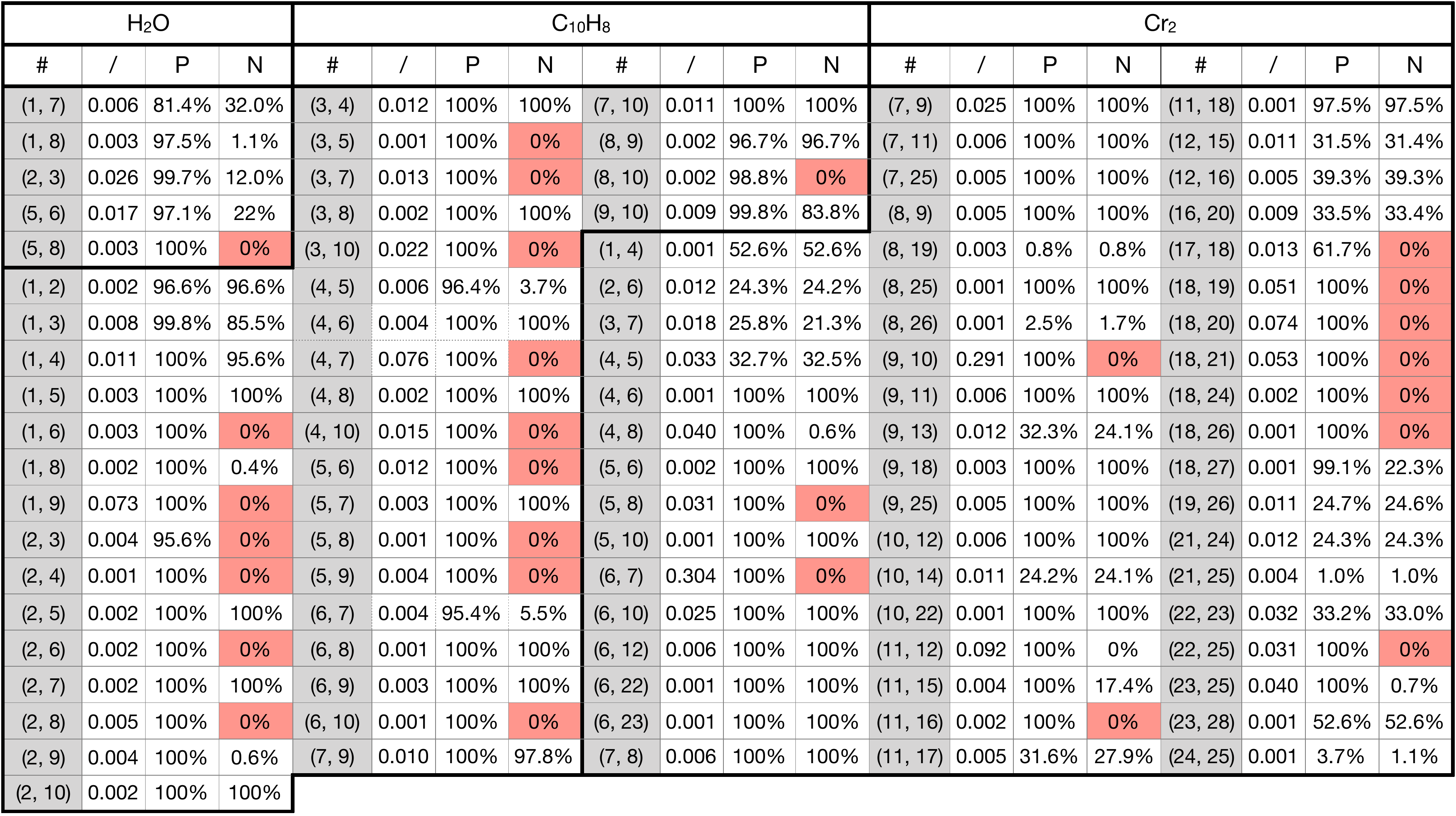}
\caption{Exact values of the orbital-orbital entanglement ($\geq0.001$) without SSR (/) and the fraction of it (in \%) remaining in the presence of P-SSR (P) and N-SSR (N), for the ground states of $\mathrm{H_2 O}$, $\mathrm{C_{10}H_8}$ and $\mathrm{Cr_2}$.}
\label{fig:2rdm}
\end{figure*}

In Figure \ref{fig:CorrWater}, \ref{fig:CorrNaph} and \ref{fig:CorrCr2_28} we present the different types of correlation of the ground state of $\mathrm{H_2O}$ constructed with 8 orbitals, $\mathrm{C_{10}H_8}$ with 10 orbitals and $\mathrm{Cr_2}$ with 28 orbitals, respectively. In Figure \ref{fig:2rdm} we list the exact value of orbital-orbital entanglement and the fraction of the entanglement (in \%) which is remaining in the presence of the P-SSR and N-SSR, respectively.

There are several important messages to get across. First of all, similar to the results for the single-orbital entanglement, the water molecule contains the weakest orbital-orbital correlation, and the chromium dimer the strongest. Most importantly, our comprehensive analysis then reveals that the quantum part of the total correlation plays only a minor role. In fact, the orbital-orbital entanglement is usually one order of magnitude smaller than the total correlation, and the molecular structure is thus dominated by classical correlation. This key result of our analysis emphasizes that the quantum mutual information \eqref{mutI} is not a suitable tool for quantifying orbital entanglement, as it leads to a gross overestimation. From a general point of view, our findings raise questions about the significance of entanglement in chemical bonding  and quantum chemistry in general.
\begin{figure}[htb]
    \centering
    \includegraphics[scale=0.29]{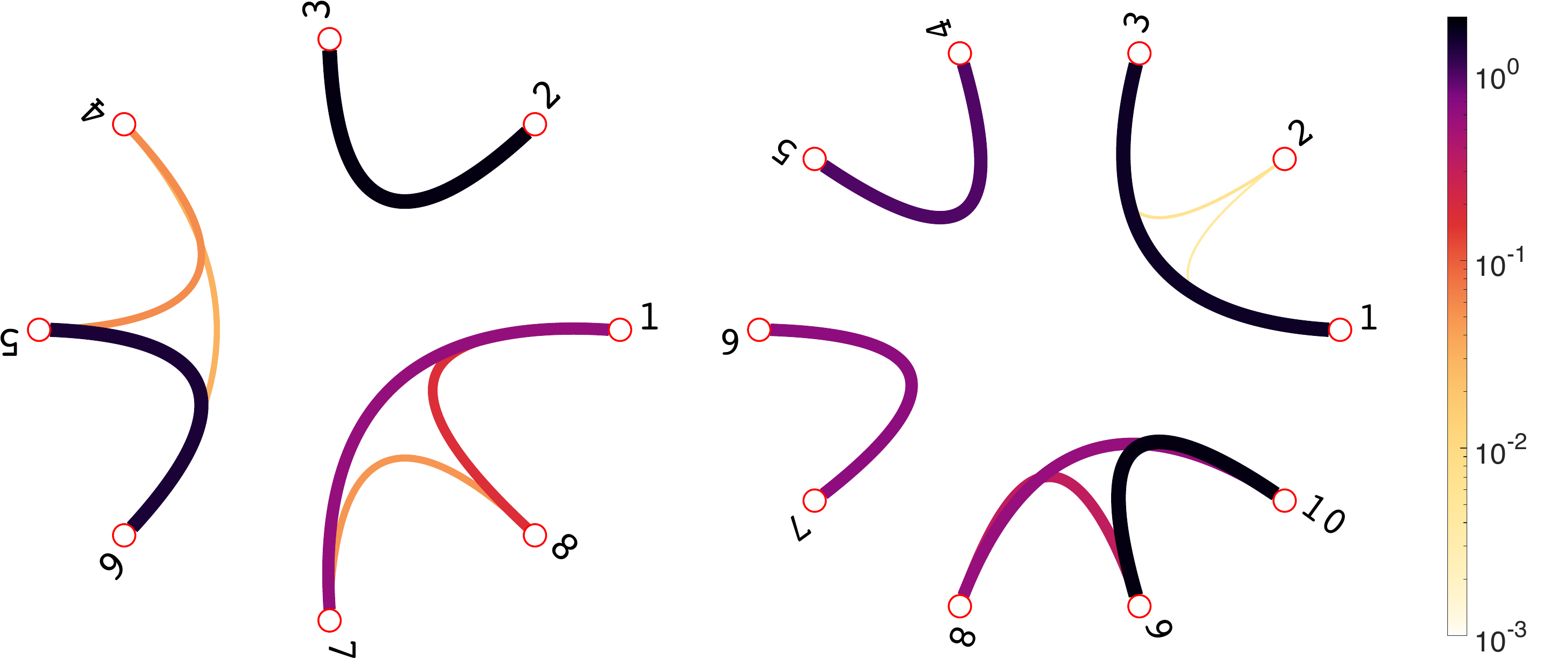}
    \caption{Orbital-orbital total correlation without SSR for $\mathrm{H_2O}$ (left) and $\mathrm{C_{10}H_8}$ (right) with the electron-electron interaction switched off.}
    \label{fig:noint}
\end{figure}

Similar to the single-orbital entanglement, SSRs also have a drastic effect on the orbital-orbital entanglement, yet in a qualitatively different way. In the molecular systems we considered, P-SSR preserved almost all of the orbital-orbital entanglement, whereas in the case of N-SSR, almost no orbital-orbital entanglement is left, and consequently almost all orbital-orbital correlation is classical. Furthermore, in some instances even the \emph{entire} orbital-orbital entanglement is destroyed by the N-SSR. Referring to Figure \ref{fig:sectors}, this indicates that most of the contribution to orbital-orbital correlation and entanglement comes from superposing $f^\dagger_{i\downarrow}f^\dagger_{i\uparrow}|{\Omega}\rangle$ and $f^\dagger_{j\downarrow}f^\dagger_{j\uparrow}|{\Omega}\rangle$, which are marked as the dark grey blocks. These states describe either empty or doubly occupied orbitals. In fact, in all three molecules, single excitations are highly suppressed in any of the molecular orbitals we consider. This is qualitatively different to the analysis of a single bond in Section \ref{sec:diss_H2} which was referring to \emph{localized atomic} orbitals, each singly occupied. In agreement with valence bonding theory, this observation confirms that two-orbital correlation and entanglement are suitable tools for describing bonding orders only if they are applied to  localized atomic orbitals.

Lastly, we would like to relate the orbital-orbital correlation pattern to the one-particle Hamiltonian $T$. Due to the point group symmetry arising from the molecular geometry, $T$ represented with respect to the active molecular orbitals is block-diagonal. This is well illustrated in  Figure \ref{fig:OneHam} not showing any coupling in $T$ between Hartree-Fock orbitals belonging to different irreducible symmetry sectors. Exploiting this structure can improve the implementation of numerical methods such as DMRG. Yet, the orbital-orbital correlation patterns inherit that structure only in case the respective ground state is weakly correlated. For the three molecules studied in our work, this is only
the case for the water molecule in agreement to its weak intrinsic correlation as quantified by \eqref{eqn:lambdadist}.
For the other two molecules the more dominant electron-electron interaction results in a major deviation of the orbital-orbital correlation patterns. To confirm these claims, we plotted the orbital-orbital total correlation similar to the top-left cells in Figure \ref{fig:CorrWater} and \ref{fig:CorrNaph}, with the same reference orbitals but with the electron-electron interaction switched off in Figure \ref{fig:noint}. We can see that the correlation patterns now match well the respective structures of the one-particle Hamiltonians in Figure \ref{fig:OneHam}. As a result, much care is needed when using the one-particle Hamiltonian to achieve a localized orbital arrangement, as the unperturbed orbital-orbital correlation and entanglement patterns might be completely scrambled by electron-electron interaction.

\section{Summary and Conclusion}\label{sec:concl}
A series of recent studies\cite{Reiher11,Reiher12,Reiher13,Reiher14,Reiher15,Ayers15a,Ayers15b,Bogus15,Reiher16,Eisert16mode,Reiher17a,Szalay17,Reiher17b,Legeza18,Legeza19} has established the quantum mutual information between orbitals as a major descriptor for electronic structure. Our work adds two important missing facets to this promising recent development:
The separation of the total correlation into classical and quantum parts and their quantification in an operational meaningful way, as required for applications in the quantum information sciences.



For this, we have first observed that the quantum mutual information is not an appropriate tool for quantifying the ``orbital entanglement''. Namely, it accounts for both quantum \emph{and} classical correlation. As part of a refined discussion of quantum information theoretical concepts, we presented a framework for separating orbital correlation into its quantum and classical parts. In particular we succeed in calculating the orbital entanglement which is typically a notoriously difficult task.

Anticipating future applications in the quantum information sciences, we quantified entanglement and correlation in an \emph{operationally meaningful} way and comprehensively explained how the formula \eqref{mutI} for the quantum mutual information then needs to be modified. For this, we introduced and explained the number parity (P-SSR) and particle number (N-SSR) superselection rule. In the form of a  communication protocol illustrated in Figure \ref{fig:QCScheme}
we provided an elementary justification of the SSRs: The violation of an SSR would make superluminal signaling possible in contradiction to the laws of relativity. However, as far as electronic structure theory is concerned, the correlation measure \eqref{mutI} has its merits as well: Since it refers to larger algebras of observables (including unphysical ones) it provides more insights into the structure of molecular ground states
than the operationally meaningful variant \eqref{eqn:SSRmeasures}.

Equipped with our distinctive measures we can eventually quantify the total correlation, entanglement and classical correlation (covering all three cases, i.e., no, P- and N-SSR) between orbitals. After an analytic illustration in Section \ref{sec:analytic}, we quantified in Section \ref{sec:results} the different correlation types exemplarily in the ground states of the water, naphthalene and dichromium molecule in a numerically exact way. Our findings as presented in Figures \ref{fig:1rdm}-\ref{fig:CorrCr2_28} reveal the following: (i) Compared to the correlation between two (orthonormalized) atomic orbitals in single bonds (order $2\ln{2}$), the correlation between most Hartree-Fock orbitals is quite small. This highlights the well-known fact that Hartree-Fock orbitals are a much better starting point for high precision ground state methods than atomic orbitals. (ii) Taking into account the important N-SSR has a drastic effect. It reduces the correlation and entanglement of one orbital with the remaining ones by about 86-96\%. The effect on the two orbital level is significant as well but varies a lot more namely between no reduction and total cancelation (see Figure \ref{fig:2rdm}). Those particular findings raise first doubts about the usefulness of molecular systems as a source for correlation and entanglement. This conclusion may change to some extent, however, if one refers to localized atomic orbitals instead of delocalized molecular orbitals. (iii) The overwhelming part of the total correlation between molecular orbitals is classical. This immediately raises questions about the role of entanglement in the description of chemical bonds and quantum chemistry in general.

Finally, we speculate that a possible comprehensive validation of the latter point (iii) in future studies could have transformative consequence:  The recent endeavor\cite{Aspuru19} to solve the electron correlation problem on a \emph{quantum} computer  would possibly loose a part of its motivation: Determining \emph{optimal} reference molecular orbitals would turn this fundamental problem almost into a classical one and in that sense reflect well the philosophy of conventional molecular bonding theory.

\begin{acknowledgements}
For the numerical calculations we used the SyTen toolkit \cite{SyTen} and for the orbital visualizations the software ChemCraft \cite{chemcraft}.
We acknowledge financial support through the J\'anos Bolyai Research Scholarship, the UKNP Bolyai+ Grant, and the NKFIH Grants No.~K124152, K124176 KH129601, K120569, and from the Hungarian Quantum Technology National Excellence Program, Project No.~2017-1.2.1-NKP-2017-00001 (S.S. and Z.Z.), from the Deutsche Forschungsgemeinschaft (DFG, German Research Foundation) under Germany's Excellence Strategy-426 EXC-2111-390814868 (U.S.) and Grant SCHI 1476/1-1 (C.S.). We also acknowledge support from the Munich Center for Quantum Science and Technology (L.D., S.M., S.D., U.S., C.S.) and the Wolfson College Oxford (C.S.).
\end{acknowledgements}

\appendix

\section{Deriving Formulas for the Single-Orbital Correlation and Entanglement} \label{app:single}
This section is devoted to deriving the formulas in Eq.~\eqref{eqn:single_SSR} for the single-orbital correlation and entanglement under P-SSR and N-SSR. In Section \ref{sec:SSRincorp} we defined the physical part of a quantum state $\rho$ under P-SSR and N-SSR using the projections
\begin{equation}
\begin{split}
\rho^\textrm{P} &= \sum_{\tau, \tau' = \textrm{odd},\textrm{even}} P_\tau \otimes P_{\tau'} \rho P_\tau \otimes P_{\tau'},
\\
\rho^\textrm{N} &= \sum_{m=0}^{\nu} \sum_{n=0}^{\nu'} P_m \otimes P_n \rho P_m \otimes P_n,
\end{split}
\end{equation}
where $\nu$ and $\nu'$ are the maximal particle numbers allowed on the local subsystems. The total correlation and entanglement available in $\rho$ given the local algebras of observables are restricted by P-SSR and N-SSR, are quantified as the total correlation and entanglement in $\rho^\textrm{P}$ and $\rho^\textrm{N}$ respectively, without the restrictions of superselection rules, according to \eqref{eqn:SSRmeasures}.

By referring to the splitting between orbital $j$ and the remaining ones, resulting in factorizing the total Fock space as $\mathcal{F}= \mathcal{F}_j \otimes \mathcal{F}_{\setminus \{j\}}$, and also assuming particle number and spin symmetries, the ground state $|\Psi\rangle$ of the total system admits the following Schmidt decomposition
\begin{eqnarray}
|\Psi\rangle &=&\sqrt{ p_1} |\Omega\rangle \otimes |\Phi_{N,M}\rangle + \sqrt{p_2} |\!\uparrow\rangle \otimes |\Phi_{N-1,M-\frac{1}{2}}\rangle
\\
&& \quad + \sqrt{p_3} |\!\downarrow\rangle \otimes |\Phi_{N-1,M+\frac{1}{2}}\rangle + \sqrt{p_4} |\! \uparrow\downarrow\rangle \otimes |\Phi_{N-2,M}\rangle. \nonumber
\end{eqnarray}
If we consider the P-SSR, the coherent terms between different local parity sectors are excluded according to Eq.~\eqref{eqn:tilde} and Figure \ref{fig:sectors}, leading to the physical state
\begin{equation}
\rho^\textrm{P} = (p_1 + p_4) |\Psi_{\textrm{even}}\rangle\langle\Psi_{\textrm{even}}| + (p_2 + p_3) |\Psi_{\textrm{odd}}\rangle \langle \Psi_{\textrm{odd}}|,
\end{equation}
where $\rho = |\Psi\rangle\langle\Psi|$ and
\begin{equation}
\begin{split}
|\Psi_{\textrm{even}}\rangle &\equiv \sqrt{\frac{p_1}{p_1+p_4}} |\Omega\rangle \otimes |\Phi_{N,M}\rangle
\\
&\quad + \sqrt{\frac{p_4}{p_1+p_4}} |\! \uparrow\downarrow\rangle \otimes |\Phi_{N-2,M}\rangle,
\\
 |\Psi_{\textrm{odd}}\rangle &\equiv \sqrt{\frac{p_2}{p_2+p_3}}  |\!\uparrow\rangle \otimes |\Phi_{N-1,M-\frac{1}{2}}\rangle  \\
 &\quad + \sqrt{\frac{p_3}{p_2+p_3}}  |\!\downarrow\rangle \otimes |\Phi_{N-1,M+\frac{1}{2}}\rangle.
 \end{split}
\end{equation}
Similarly for N-SSR, the physical state is
\begin{equation}
\rho^\textrm{N} = p_1 |\Psi_0\rangle\langle\Psi_0| + (p_2+p_3) |\Psi_1\rangle\langle\Psi_1| + p_4 |\Psi_2\rangle\langle\Psi_2|,
\end{equation}
where
\begin{equation}
\begin{split}
|\Psi_0\rangle &\equiv  |\Omega\rangle \otimes |\Phi_{N,M}\rangle ,
\\
|\Psi_1\rangle & \equiv |\Psi_\textrm{odd}\rangle,
\\
|\Psi_2\rangle & \equiv |\! \uparrow\downarrow\rangle \otimes |\Phi_{N-2,M}\rangle.
\end{split}
\end{equation}
To calculate the total correlation, we first determine the spectra of the respective reduced density matrices $\rho^\textrm{P}_j$, $\rho^\textrm{P}_{\setminus \{j\}}$ and $\rho^\textrm{N}_j$, $\rho^\textrm{N}_{\setminus \{j\}}$. Due to the highly symmetric total state, all four matrices are isospectral as $\rho_1$ in \eqref{eqn:1rdm}. Using the definition of the quantum mutual information in \eqref{eqn:mutual_info} we obtain
\begin{equation}
\begin{split}
I(\rho^\textrm{P}) &= (p_1 + p_4) \ln(p_1 + p_4) + (p_2 + p_3)\ln(p_2+p_3)
\\
 -& 2(p_1 \ln(p_1) + p_2 \ln(p_2) + p_3 \ln(p_3) + p_4 \ln(p_4)),
 \\
 I(\rho^\textrm{N}) & = p_1 \ln(p_1) + (p_2 + p_3)\ln(p_2+p_3) + p_4 \ln(p_4)
\\
 -& 2(p_1 \ln(p_1) + p_2 \ln(p_2) + p_3 \ln(p_3) + p_4 \ln(p_4)).
\end{split}
\end{equation}
For the single-orbital entanglement, we use the result obtained in Ref.~\onlinecite{LexinQIT}, which allows us, given certain criteria are met, to separate the single-orbital entanglement into the entanglement of its pure state decomposition,
\begin{equation}
\begin{split}
E(\rho^\textrm{P}) &= (p_1+p_4) E(|\Phi_\textrm{even}\rangle\langle\Phi_\textrm{even}|)
\\
&\quad \quad +  (p_2+p_3) E(|\Phi_\textrm{odd}\rangle\langle\Phi_\textrm{odd}|),
\\
E(\rho^\textrm{N}) & = (p_2+p_3) E(|\Phi_\textrm{odd}\rangle\langle\Phi_\textrm{odd}|).
\end{split}
\end{equation}
Using the von Neumann entropy in \eqref{eqn:ententr} as the entanglement measure for pure states, the single-orbital entanglement in the presence of P-SSR and N-SSR is also determined solely by the spectrum of the one-orbital reduced density matrix,
\begin{equation}
\begin{split}
E(\rho^\textrm{P}) &= (p_1 + p_4) \ln(p_1 + p_4) + (p_2 + p_3)\ln(p_2+p_3)
\\
& - p_1 \ln(p_1) - p_2 \ln(p_2) - p_3 \ln(p_3) - p_4 \ln(p_4).
\\
E(\rho^\textrm{N}) &= (p_2 + p_3) \ln(p_2 + p_3) - p_2 \ln(p_2) - p_3 \ln(p_3).
\end{split}
\end{equation}

\vspace{5cm}

\bibliographystyle{achemso}
\bibliography{qchem-v6JCTC}

\newpage

\section*{}
\vspace{2cm}
\section*{Table of Contents (TOC) graphic}
\begin{figure}[htb]
    \centering
    \includegraphics[scale=0.24]{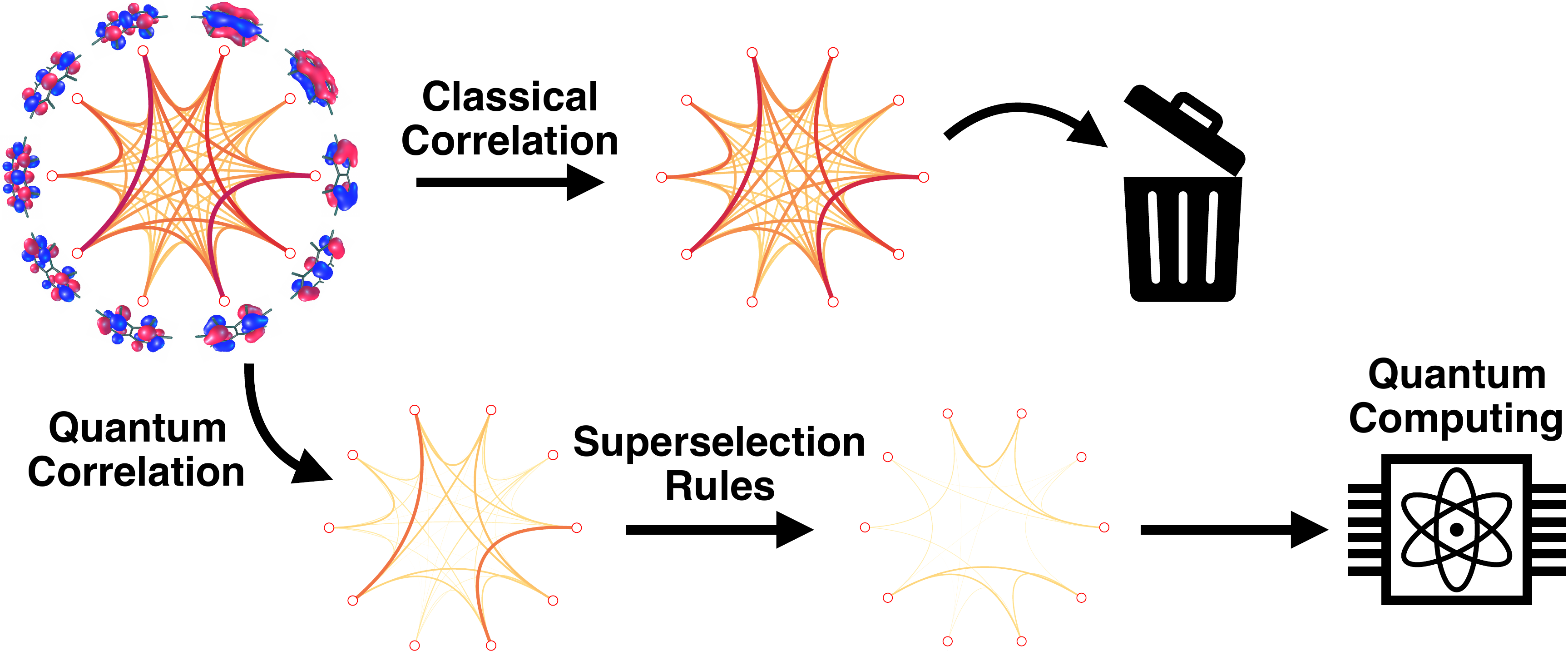}
    \caption{Our work comprehensively explains how the correlation in a molecule separates into the rather useless classical correlations and the quantum correlations. Surprisingly, the latter are found to play only a minor role in chemical bonding, partly due to a fundamental rule of relativity. This raises questions about the role of entanglement in quantum chemistry and the usefulness of molecules for quantum computing.}
\end{figure}

\end{document}